\documentclass[journal]{IEEEtran}
\usepackage{algorithm}
\usepackage{algpseudocode}
\usepackage{fancyhdr}
\usepackage{textcomp}
\usepackage{amsmath}
\usepackage{subfigure}
\usepackage{amssymb,amsfonts}
\usepackage{graphicx}
\usepackage{textcomp}
\usepackage{xcolor}
\usepackage{multicol}
\usepackage{multirow}
\usepackage{blindtext}
\usepackage{adjustbox}

\usepackage{balance}
\usepackage{siunitx}
\usepackage{makecell}

\usepackage{array}
\algnewcommand{\Initialize}[1]{%
  \State \textbf{Initialize:}
\parbox[t]{.8\linewidth}{\raggedright #1}
}
\algnewcommand{\Goto}{\textbf{go to}}%

\def\BibTeX{{\rm B\kern-.05em{\sc i\kern-.025em b}\kern-.08em
    T\kern-.1667em\lower.7ex\hbox{E}\kern-.125emX}}

\usepackage{soul}    
\usepackage{xcolor}
\setstcolor{red}     
\usepackage{graphicx}
\usepackage{ulem}
\usepackage{booktabs}
\begin{document}

\title{Multi-Modal Data-Enhanced Foundation Models \\for Prediction and Control in Wireless Networks: \\A Survey}

\author{Han~Zhang,
        Mohammad Farzanullah,
        Mohammad Ghassemi,
        Akram Bin Sediq,
        Ali Afana,
        and~Melike~Erol-Kantarci,~\IEEEmembership{Fellow, IEEE}
\thanks{Han Zhang, Mohammad Farzanullah, Mohammad Ghassemi and Melike Erol-Kantarci are with the School of Electrical Engineering and Computer Science, University of Ottawa, Ottawa, ON K1N 6N5, Canada (e-mail: hzhan363@uottawa.ca; mfarz086@uottawa.ca; mghas017@uottawa.ca; melike.erolkantarci@uottawa.ca).}
\thanks{Akram Bin Sediq and Ali Afana are with Ericsson, Ottawa, K2K 2V6, Canada (e-mail:
akram.bin.sediq@ericsson.com; ali.afana@ericsson.com)
}}


\maketitle

\begin{abstract}
Foundation models (FMs) are recognized as a transformative breakthrough that has started to reshape the future of artificial intelligence (AI) across both academia and industry. The integration of FMs into wireless networks is expected to enable the development of general-purpose AI agents capable of handling diverse network management requests and highly complex wireless-related tasks involving multi-modal data. Inspired by these ideas, this work discusses the utilization of
FMs, especially multi-modal FMs in wireless networks. We focus on two important types of tasks in wireless
network management: prediction tasks and control tasks. In particular, we first discuss FMs-enabled multi-modal contextual information understanding in wireless networks. Then, we explain how FMs can be applied to prediction and control tasks, respectively. Following this, we introduce the development of wireless-specific FMs from two perspectives: available datasets for development and the methodologies used. Finally, we conclude with a discussion of the challenges and future directions for FM-enhanced wireless networks.
\end{abstract}

\begin{IEEEkeywords}
Generative foundation models, multi-modal foundation models, Generative AI, large language models, wireless networks, traffic prediction, B5G 
\end{IEEEkeywords}

\IEEEpeerreviewmaketitle

\section{Introduction}
%
%
%
%
\IEEEPARstart{T}{he} rise of foundation models (FMs) marks a significant breakthrough that is shaping the artificial intelligence (AI) field. FMs refer to large-scale machine learning (ML) models trained on extensive datasets that can be adapted and fine-tuned for various applications and downstream tasks \cite{ArtificialIntelligenceFoundation}. Given their high generality and adaptability, there is a growing trend of using open-source FMs as the backbone of scenario-oriented AI applications. This approach effectively reduces engineering costs and minimizes the need for extensive involvement in model design and training. As FMs grow in popularity, the AI landscape is expected to shift from developing custom-built AI systems to creating general-purpose, deployment-ready AI applications \cite{FoundationModelsANewParadigm}\cite{TowardsArtificialGeneral}. 

The emergence of the transformer architecture has dramatically accelerated the progress of FMs \cite{AComprehensiveSurvey}. The transformer is a neural network architecture based on the multi-head self-attention mechanism \cite{AttentionIsAllYouNeed}. It typically consists of alternating attention and feed-forward layers that perform repeated transformations on the vector representations and extract latent information. The introduction of the attention mechanism in the transformer architecture allows elements in the input sequence to be weighted based on their importance and enables the model to capture long-range dependencies. Benefiting from this architecture, transformers exhibit scaling laws \cite{ScalingLawsForNeural}, showing that the model's performance predictably increases when the model size, dataset size, and compute budget increase. The transformer’s scaling law ensures its ability to scale effectively over ultra-large datasets. This led to the idea of using large-scale transformer-based models trained on massive datasets to gain emergent capabilities and generalization capabilities on a series of downstream tasks, which motivated the emergence and popularity of FMs. Compared with traditional AI techniques, FMs have two unique characteristics: emergence and homogenization \cite{OnTheOpportunities}. Emergence means that the model behavior is induced from training on extensive data rather than being explicitly constructed.  Homogenization means the FM is a single generic learning algorithm that can handle a wide range of applications. These two characteristics enhance the usability of FMs across diverse data types and domains.

\begin{table*}[]
\centering
\renewcommand\arraystretch{1.2}
\caption{A comparison between the definition and capabilities of LLM, Generative AI, and FM.}
\label{table-introduction}
\begin{tabular}{|c|c|c|c|c|}
\hline
Concepts                                                & Definition                                                                                                                                                                                                                            & Capabilities                                                                                                                            & Examples                                                                                          \\ \hline
LLM                                                     & \begin{tabular}[c]{@{}c@{}}Transformer-based ML \\ models that can comprehend and \\ generate human language text \cite{ASurveyOfLargeLanguageModels}.\end{tabular}                                                                                         & \begin{tabular}[c]{@{}c@{}}Multi-modal information understanding\\  and generation, emergent capabilities,\\  reasoning and so on\end{tabular} & \begin{tabular}[c]{@{}c@{}}Generative pre-training transformer (GPT) \cite{GPT4MODEL,GPT2MODEL}, \\Pathways language model (PaLM) \cite{PALMMODEL}, \\ Large language model Meta AI (LLAMA) \cite{LLAMAMODEL}\end{tabular}                                       \\ \hline
\begin{tabular}[c]{@{}c@{}}Generative\\ AI\end{tabular} & \begin{tabular}[c]{@{}c@{}}A class of algorithms and models\\ within AI and NLP designed to generate\\  new, previously unseen data rather than\\ distinguishing between existing categories\\ as discriminative models \cite{RecentAdvnces}.\end{tabular} & \begin{tabular}[c]{@{}c@{}}Multi-model content generation,\\ fraud detection, recommendations\end{tabular}                              & \begin{tabular}[c]{@{}c@{}}Generative adversarial network (GAN) \cite{GAN}, \\Variational autoencoder (VAE) \cite{VAE}, \\ Autoregressive Models, \\ Diffusion Models\end{tabular} \\ \hline
FM                                                      & \begin{tabular}[c]{@{}c@{}}A series of models trained on broad data\\  that can be adapted to a wide range\\  of downstream tasks, making them different\\ from task-agnostic models \cite{OnTheOpportunities}.\end{tabular}                                    & \begin{tabular}[c]{@{}c@{}}Emergence, homogenization, \\ and domain adaption,\\ serves as the common basis\end{tabular}                 & \begin{tabular}[c]{@{}c@{}}GPT \cite{GPT4MODEL,GPT2MODEL}, Bidirectional encoder \\representations from transformers (BERT) \cite{BERTMODEL}, \\DALL-E \cite{DALLEMODEL}\end{tabular}                                                                  \\ \hline
\end{tabular}
\end{table*}

In recent years, various AI techniques have been widely employed in all aspects of beyond fifth-generation (B5G) wireless communications, including optimizing network performance, improving network security, and automating the manual processes involved in managing wireless networks \cite{ExploringThePotentialOfAI}. These techniques possess the capabilities to handle the complexity of the network architecture and the growing demand for communications \cite{AIEnabledFuture}. However, most of the current widely used AI-driven network management applications are based on small-scale custom-built AI models.

While existing small-scale, custom-built AI techniques have significantly accelerated the evolution of wireless networks, several opportunities remain to be seized. A primary concern is that custom-built AI techniques are usually only applicable to a single use case or a limited number of specific use cases. Once the scenario changes, the model has to be rebuilt and trained again. Future intelligent communications are expected to consider not only how AI techniques can be used in specific network optimization problems or scenarios. They are expected to also incorporate AI into all aspects of the wireless communication system and propose a generic approach to designing AI-driven wireless communications and realizing the vision of intrinsic AI \cite{TelecomsArtificialGeneralIntelligence}. Therefore, there is a need to build general-purpose AI models that are applicable to broader types of use cases in wireless networks \cite{ArtificialGeneralIntelligenceNativeWireless}.

The second issue is that the complexity of network management tasks in wireless communication has been continuously increasing. The next-generation network represents a highly complex and heterogeneous structure with diversified quality of service (QoS) requirements. Subscriber-centric service is becoming a key enabler to the B5G success, which means service providers are expected to understand customers' needs and provide tailored network capabilities and network characteristics according to the requirements of the customers \cite{6GNetworkBusinessSupport}. This makes some of the network management tasks beyond the capabilities of regular small-scale AI models \cite{BigAIModelsFor6G}. 

\begin{figure*}[!t]
\centering
\includegraphics[width=5.3in]{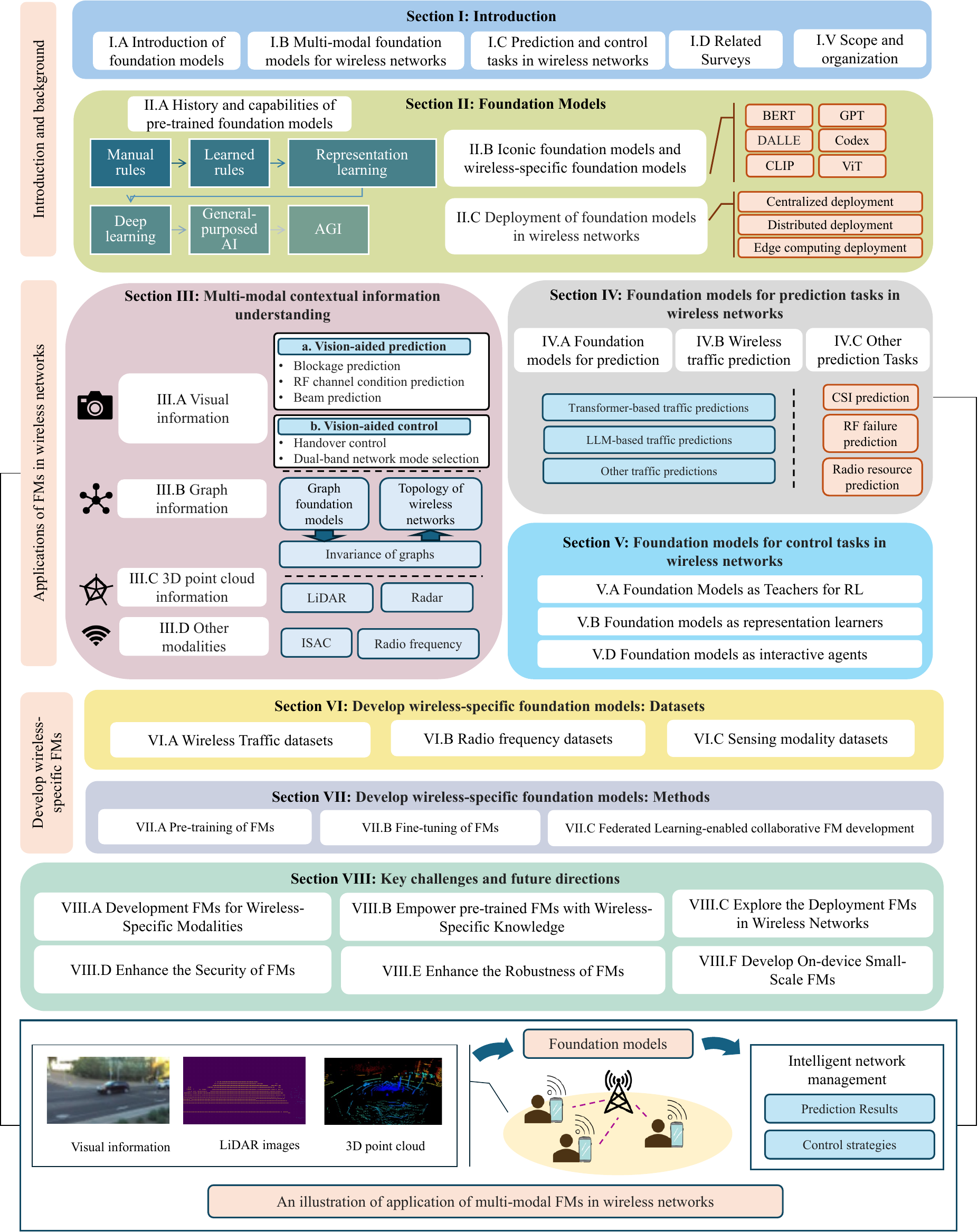}
\caption{Organization and key topics covered in this work and an illustration of applications of FMs in wireless networks.}
\label{organization}
\vspace{-10pt}
\end{figure*}

Another concern is that contextual information in wireless communication is becoming increasingly diverse. With the development of different sensors on base stations (BSs) and mobile devices, various types of sensory data with different operating ranges and properties have been collected and included in wireless network management \cite{IntelligentMultiModalSensing}. These data include, but are not limited to, radio frequency (RF) data, radio detection and ranging (RADAR) data, light detection and ranging (LiDAR) data, red-green-blue (RGB) camera data, and the global positioning system (GPS) \cite{alkhateeb2023deepsense}. 

Multi-modal sensing can be leveraged for various channel-related downstream tasks, including channel estimation, blockage prediction, beam selection, and handover control. As a result, there is a chance to leverage multi-modal learning to improve the performance of communication systems in the B5G era. However, training multi-modal models incurs a high computational cost and typically requires a large and diverse dataset. Training custom-built AI models from scratch for each use case will impose a heavy computational burden on service providers and exacerbate data scarcity issues. More importantly, the AI models being trained across different tasks tend to have commonalities when the input information is of the same modalities. This underscores the possibility of fine-tuning a single underlying base model for multiple downstream tasks.

The discussions above highlight the need for utilizing FMs, especially multi-modal FMs in wireless networks. There are two important types of tasks to be handled in wireless network management, prediction tasks and control tasks. Prediction refers to the tasks that use data-driven techniques to forecast future information on wireless networks. The predicted network conditions, user behaviors, and resource demands can be used to improve the network performance \cite{ASurveyOfOnline}. Control tasks refer to the process that dynamically manages network resources and selects optimal network configurations to achieve wireless network goals, including more reliable connectivity, higher QoS, and higher energy efficiency \cite{ApplicationsOfDeepReinforcementLearning}. FMs can be applied to both types of tasks to develop intelligent prediction and decision-making schemes for self-organized wireless network management.

\vspace{-10pt}
\subsection{Related Surveys}

The discussion of FMs in the literature often intersects with two closely related concepts: large language models (LLMs) and generative AI. Table I compares these three concepts based on their definition and capabilities and distinguish the differences between them. A similarity between the three concepts is that the transformer structure plays an important role in all three types of models. Specifically, LLMs can be regarded as a special subset of generative FMs capable of performing a variety of natural language processing (NLP) tasks \cite{EvolutionAndProspectsOfFoundation}. 

However, an overemphasis on LLMs can obscure the broader potential of other FM architectures, which possess unique strengths for wireless communications. FMs are not limited to text; they also include powerful architectures for other modalities, such as Vision Transformers (ViT) for visual data \cite{ViTMODEL}, and Graph Neural Networks (GNNs) increasingly adapted as FMs for topological data \cite{shen2022graph}. This architectural diversity is critical for wireless applications. For instance, vision FMs can analyze camera data for LoS beam selection \cite{jiang2022camera} or process spectrum spectrograms for interference identification \cite{huo2024recent}. Similarly, GNN-based FMs are inherently suited for modeling network topology, managing radio resources, and optimizing routing paths. A balanced perspective, which this paper aims to provide, must therefore consider this full range of FM architectures.

In the literature, a few recent works have made comprehensive reviews of LLMs \cite{ASurveyOfLargeLanguageModels, LargeLanguageModelsAComprehensive, ASurveyOnMultimodalLarge, ASurveyOnEvaluationOfLarge}. These papers give detailed introductions about LLMs and multi-modal LLMs, covering aspects including the history and evolution of LLMs, publicly available resources of LLMs, the back-end structure of LLMs, the training and fine-tuning methods of LLMs, and the evaluation of LLMs. In addition, some papers discussed how to apply LLMs to wireless communication. \cite{PushingLargeLanguageModels} explores the potential of deploying LLMs at the sixth generation (6G) edge and identifies the critical challenges. \cite{LargeLanguageModelsIn6GSecurity} discusses the potential adversaries of LLM-based 6G networks and explores how to connect LLMs with blockchain technology. \cite{LargeLanguageModelForTelecommunications} presents LLM fundamentals and introduces LLM-enabled key techniques and telecom applications. These works are related to our research, however, they focus more on LLMs. Other FMs, especially FMs for multi-modal data processing, have not been studied thoroughly.

Similarly, generative AI has been discussed in some existing surveys \cite{AComprehensiveSurveyOfAIGenerated, ASurveyOfGenerativeArtificial, ThePowerOfGenerativeAi}, which have covered the basic components and recent advances of generative AI. Specifically, some works studied the application of generative AI in wireless communications. For instance, \cite{GenerativeAIInMobileNetworks} studies how to apply generative AI to mobile communication networks. \cite{GenerativeAIForSecurePhysicalLayer} provides an extensive survey on the various applications of generative AI in enhancing security within the physical layer of communication networks. Although these works are related to our topic, these papers talk more about the data generation and distribution modeling capabilities of generative AI models, while this work stresses more on how to leverage FMs for prediction and control in wireless networks.

The applications, capabilities, and potential challenges of FMs have been discussed thoroughly in review papers \cite{ArtificialIntelligenceFoundation, FoundationModelsANewParadigm, TowardsArtificialGeneral, AComprehensiveSurvey, ASurveyOfHallucinationIn, FewShotAdaptationOfMultiModal}. In the wireless networks context, the closest survey papers to this work are \cite{TelecomFoundationModels, AIFoundationModelsInRemoteSensing} and \cite{BigAIModelsFor6G}. \cite{TelecomFoundationModels} investigates the potential opportunities of using FMs to shape the future of telecom technologies and standards. \cite{AIFoundationModelsInRemoteSensing} provides a comprehensive survey of FMs in the remote sensing domain and categorizes models based on their applications. \cite{BigAIModelsFor6G} gives some in-depth prospects on the demand, design, and deployment aspects of the wireless big AI models. In contrast to these works, this study has a distinct focus, concentrating specifically on multi-modal data and two primary types of tasks in wireless networks: prediction and control tasks. Moreover, compared with \cite{TelecomFoundationModels} and \cite{AIFoundationModelsInRemoteSensing}, this work conducts a more in-depth and comprehensive survey and offers a richer level of detail. While \cite{TelecomFoundationModels} and \cite{AIFoundationModelsInRemoteSensing} focus more on conceptual insights and visionary perspectives, this work thoroughly examines existing studies and current literature for detailed understanding.

\subsection{Contributions and Organization}

Although FMs have demonstrated exceptional generalization across diverse tasks and have been used to solve complicated problems in various domains, their full potential remains under-explored. In particular, there is a lack of comprehensive studies on the current and prospective applications of FMs, especially multi-modal FMs, in AI-driven wireless network management. In this survey paper, we explore the application of FMs for processing multi-modal data in wireless networks. Specifically, we examine their potential usage in two key types of wireless network management tasks: prediction tasks and control tasks. This survey lays the foundation for researchers to advance AI-enabled applications in the era of multi-modal, data-enhanced network management, opening up new possibilities for integrating FMs with traditional AI-driven wireless networks. The contributions of our survey are listed below:

1) We provide a comprehensive survey on the applications of FMs, with a particular emphasis on multi-modal FMs in wireless networks. We discuss the evolution, capabilities, prominent examples, and deployment of FMs and detail how these models can be effectively utilized to process contextual data from diverse modalities for network management.

2) We focus on two key categories of tasks in wireless networks, prediction tasks and control tasks, and we explore how FMs can be applied to these tasks. For prediction tasks, we study both FMs for wireless traffic prediction and FMs for other prediction tasks such as channel state information (CSI) prediction, RF failure prediction, and blockage prediction. For control tasks, we discuss how FMs can facilitate the exploration strategies of control agents, extract latent representations from their observations, and function as interactive agents in control scenarios.

3) We conduct a detailed survey on how to build wireless-specific FMs from two key perspectives: datasets and methodologies. First, we summarize the currently available wireless communication datasets suitable for FM development. Then, we explore the methods for fine-tuning and training FMs to adapt to wireless networks.

Fig. \ref{organization} presents the organization of this work. As shown in Fig. \ref{organization}, the rest of this work is structured as follows: In section \ref{s2}, a comprehensive survey of FMs is given covering the history and capabilities of pre-trained FMs, a review of well-known FMs, as well as wireless specific FMs, and the deployment of FMs in wireless networks. In Section \ref{s3}, we explore the application of FMs for understanding multi-modal contextual information. This includes various modalities, such as visual information, graph information, three-dimensional (3D) point cloud information, and other modalities relevant to wireless networks. Section \ref{s4} and Section \ref{s5} focus on the applications of FMs in prediction tasks and control tasks in wireless networks. In Section \ref{s6} and Section \ref{s7}, we discuss how to develop wireless-specific FMs. Section \ref{s6} covers single-modal and multi-modal wireless communication datasets suitable for FM development, and Section \ref{s7} focuses on the methods of training and task adaptation of FMs in the wireless communication domain. Finally, in Section \ref{s8}, we discuss the key challenges and potential future directions of applying FMs in wireless networks. In Section \ref{s9}, we conclude the paper.

\section{Background on Foundation Models}
\label{s2}

In this section, a step-by-step survey is presented to make a detailed introduction to FM fundamentals. This includes the evolution and capabilities of FMs, the taxonomy of FMs, well-known FMs, wireless-specific FMs, and the deployment of FMs in wireless networks. 

\vspace{-5pt}
\subsection{Evolution and Capabilities of FMs: From Representation Learning to LLMs}

\begin{figure}[!t]
\centering
\includegraphics[width=3.2in]{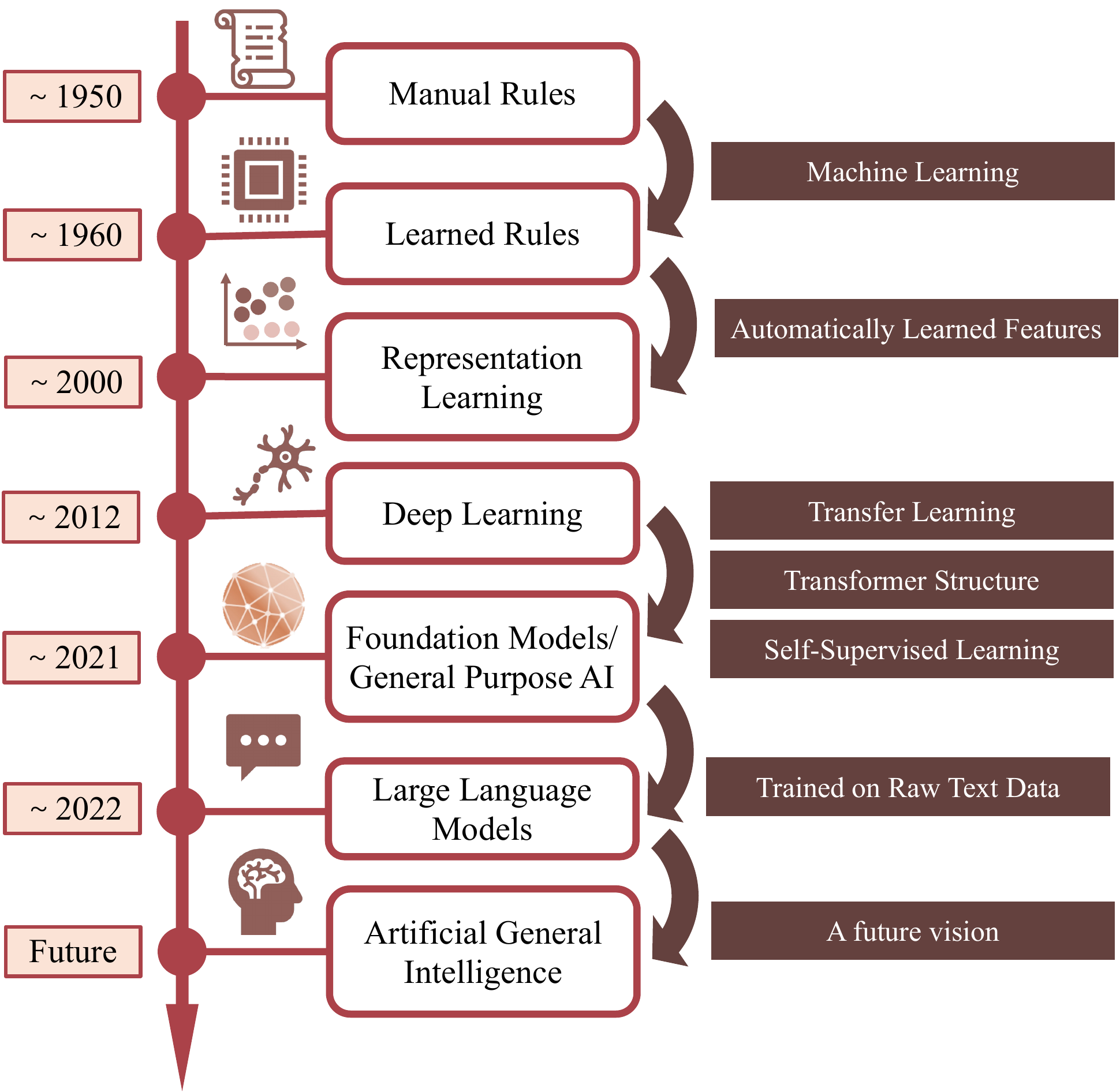}
\caption{The evolution of FMs.}
\label{Evolution}
\vspace{-10pt}
\end{figure}

The remarkable adaptability of FMs to diverse tasks stems from their ability to encapsulate versatile rules applicable across various application scenarios. By tracing the evolution of methods for extracting general rules from data, we can gain deeper insights into the development and purpose of FMs \cite{FoundationModelsANewParadigm}.

Fig. \ref{Evolution} illustrates the evolution of FMs. Before the advent of ML techniques, machines relied on pre-written rules to solve problems and make decisions. These rules were usually manually constructed by experts, leveraging their domain knowledge, and explicitly programmed into systems. With the introduction of rule-based ML techniques such as decision trees, algorithms began to extract useful rules in a supervised manner using labeled data. However, in the early stages of ML, while rules were learned automatically, the features used by the algorithms were still manually designed by human experts. To overcome this limitation, representation learning techniques such as principal component analysis (PCA) were introduced, enabling machines to automatically extract meaningful patterns from raw data and learn task-specific representations \cite{AnOverviewOnDataRepresentation}.

In recent years, deep learning (DL) was proposed to build ML models in a modular, flexible way by stacking various layers of neurons on top of each other \cite{MachinLearningAndDeepLearning}. DL enables machines to learn features directly from data, reducing the need for feature engineering. The layering structure also facilitates the scalability of DL models and allows models to generalize effectively on larger datasets. DL models are typically trained in a supervised manner using labeled data, which makes them susceptible to data scarcity and privacy challenges. To mitigate these concerns, transfer learning was proposed, allowing models to be pre-trained and then fine-tuned for specific downstream tasks in related domains \cite{ASurveyOfTransferLearning}. This advancement paved the way for the emergence of FMs.

The term FM was first formally proposed in late 2021 by a group of research associates of Stanford's Human-Centered AI. They defined FMs as any model that is trained on broad data that can be adapted to a wide range of downstream tasks \cite{OnTheOpportunities}. Meanwhile, the transformer architecture was developed with scaling laws \cite{ScalingLawsForNeural}, enabling models to efficiently handle larger datasets and continue scaling in size and complexity. In principle, the FM can be built on any model architecture. However, transformer architecture has become a dominant choice for FMs due to its scalability and ability to capture long-range dependencies. As a result, many FMs today are based on the transformer structure.

In addition to the above, self-supervised learning has been proposed as an effective training method to address the scarcity of labeled data \cite{BeyondSupervisedTheRise}. In this approach, the pre-training task is derived from task-agnostic data with self-generated labels. While unlabeled raw text data is abundant and widely available, which makes it particularly well-suited for self-supervised learning for large FMs \cite{PretrainedLanguageModelsInBiomedical, ArtificialIntelligenceBasedMedicalDataMining}, this is not the only viable modality. The same principles apply to other broad datasets, such as the vast amount of visual and sensor data used to train ViTs and other sensory-based models. This parallel evolution is crucial for wireless applications, where non-textual data is often paramount. For example, ViTs can be trained on environmental data to perform tasks like line-of-sight (LoS) blockage prediction in 6G vehicular networks \cite{Gharsallah2024ViT}. This broader context has led to the development and popularization of LLMs as a prominent, but not exclusive, special type of FM \cite{Liu2024Survey}.

Given the extensive and varied data on which FMs are trained, they have acquired cross-domain knowledge and exhibited emergent capabilities specific to their training modality. For instance, Some LLMs, such as GPTs, are pre-trained for general-purpose language use \cite{Bhardwaj2023Comprehensive}. These text-based models can function as general-purpose interfaces \cite{Yusuf2023Language} or interactive agents \cite{wang2024survey} in language-centric tasks. Concurrently, vision-based FMs (like ViTs) demonstrate strong capabilities in spatial and environmental reasoning, which are strengths directly applicable to wireless network challenges such as resource allocation, signal propagation modeling, and blockage prediction \cite{Gharsallah2024ViT}. Other multi-modal FMs aim to fuse these domains, processing text, images, and other signals simultaneously. The emergence of these powerful and general-purpose AI models is regarded as a potential spark for Artificial General Intelligence (AGI) \cite{Bubeck2023Sparks}.

Emergence and homogenization are two major characteristics of FMs \cite{OnTheOpportunities}. Emergence means that the model behavior is induced from training on diverse and extensive data, rather than being explicitly constructed. It makes FMs adaptable task-solvers that can perform tasks beyond the scope of training objectives. Homogenization means the FM is a single generic learning algorithm that can handle a wide range of applications. The same model can be repeatedly used for fine-tuning and training for downstream tasks, as they are effective across many tasks and too cost-intensive to train individually.

Given the emergence characteristic, some FMs exhibit numerous important emergent capabilities that are absent in smaller AI models. Among these, two emergent capabilities are widely recognized and sometimes regarded as the foundation for other emergent capabilities \cite{AreEmergentAbilitiesIn, ASurveyOfReasoning}. The first prominent emergent capability is few-shot prompting \cite{wei2022emergent}. Few-shot prompting means that pre-trained large models can perform tasks by learning from a few examples provided in the input without further training or parameter updates \cite{LanguageModelsAreFew-shotLearners}. This phenomenon forms the basis for in-context learning (ICL) \cite{ASurveyOnInContextLearning}, enabling FMs to generalize dynamically to new tasks. This method has been demonstrated as effective in various FMs, including LLMs, diffusion models \cite{wang2023incontext}, and large multi-modal FMs \cite{jiang2024manyshot}. 

Another important emergent capability is reasoning, which refers to the ability of FMs, especially LLMs, to analyze, process, and conclude from the input information through logical analysis, analogical analysis, and causal analysis \cite{webb2022emergent, ASurveyOfReasoning, DoesReasoningEmerge}. The reasoning capability is the foundation for techniques such as chain of thoughts (CoT) prompting that teaches LLMs to reason through a sequence of smaller steps.

\begin{table*}[!t]
\centering
\caption{Well-known FMs in various application domains.}
\renewcommand\arraystretch{1.2}
\label{table-FMs}
\resizebox{2\columnwidth}{!}{%
\begin{tabular}{|c|c|c|c|c|c|c|}
\hline
Model Name &
  \begin{tabular}[c]{@{}c@{}}Key Model \\ Architecture\end{tabular} &
  Model Size &
  \begin{tabular}[c]{@{}c@{}}Application\\ Domain\end{tabular} &
  Functionality &
  \begin{tabular}[c]{@{}c@{}}Year \\ Realeased\end{tabular} &
  \begin{tabular}[c]{@{}c@{}}Open \\ Source\end{tabular} \\ \hline
BERT \cite{BERTMODEL} &
  \begin{tabular}[c]{@{}c@{}}Multilayer bidirectional \\ transformer encoder\end{tabular} &
  \begin{tabular}[c]{@{}c@{}}42M/108M/\\ 334M\end{tabular} &
  \multirow{2}{*}{Language} &
  Natural language understanding &
  2018 &
  Yes \\ \cline{1-3} \cline{5-7} 
T5 \cite{T5MODEL} &
  \begin{tabular}[c]{@{}c@{}}Encoder-decoder \\ transformer\end{tabular} &
  \begin{tabular}[c]{@{}c@{}}60M/220M/\\ 770M/3B/ 11B\end{tabular} &
   &
  \begin{tabular}[c]{@{}c@{}}Perform NLP tasks in a text-to-text \\ form, such as translation, \\ summarization, and question-answering\end{tabular} &
  2019 &
  Yes \\ \hline
Vision Transformer \cite{ViTMODEL} &
  Transformer encoder &
  \begin{tabular}[c]{@{}c@{}}86M/307M\\ /632M\end{tabular} &
  \multirow{5}{*}{Vision} &
  Image classification &
  2020 &
  Yes \\ \cline{1-3} \cline{5-7} 
CLIP \cite{radford2021learning} &
  \begin{tabular}[c]{@{}c@{}}ResNet-based image encoder\\  and transformer-based \\ text encoder\end{tabular} &
  152M &
   &
  \begin{tabular}[c]{@{}c@{}}Connect textual and visual information \\ and perform vision-language tasks\\  such as image classification\end{tabular} &
  2021 &
  Yes \\ \cline{1-3} \cline{5-7} 
DALL-E \cite{DALLEMODEL} &
  Decoder-only transformer &
  \begin{tabular}[c]{@{}c@{}}Not publicly \\ disclosed\end{tabular} &
   &
  \begin{tabular}[c]{@{}c@{}}Generate images from textual \\ descriptions\end{tabular} &
  2021 &
  \begin{tabular}[c]{@{}c@{}}Not officially \\ open-source\end{tabular} \\ \cline{1-3} \cline{5-7} 
DALL-E 2 \cite{DALLE2MODEL} &
  \begin{tabular}[c]{@{}c@{}}Diffusion-based decoder\\  with CLIP embeddings\end{tabular} &
  \begin{tabular}[c]{@{}c@{}}Not publicly \\ disclosed\end{tabular} &
   &
  Text-conditional image generation &
  2022 &
  No \\ \cline{1-3} \cline{5-7} 
Stable Diffusion \cite{STABLEDIFFUSION} &
  VAE and U-Net decoder &
  890M &
   &
  Generate images from text prompts &
  2022 &
  Yes \\ \hline
Codex \cite{CODEXMODEL} &
  Decoder-only transformer &
  175B &
  Code &
  Code generation and assistance &
  2021 &
  No \\ \hline
\begin{tabular}[c]{@{}c@{}}NVIDIA's Isaac\\  GR00T N1 \cite{NVIDIAISSAC}\end{tabular} &
  \begin{tabular}[c]{@{}c@{}}A dual-system\\  architecture inspired\\  by human cognition\end{tabular} &
  \begin{tabular}[c]{@{}c@{}}Not publicly\\ disclosed\end{tabular} &
  Robotics &
  \begin{tabular}[c]{@{}c@{}}Generalized humanoid\\  robot reasoning and skills\end{tabular} &
  2025 &
  Yes \\ \hline
GPT-3 \cite{LanguageModelsAreFew-shotLearners} &
  Decoder-only transformer &
  175B &
  \multirow{5}{*}{\begin{tabular}[c]{@{}c@{}}General\\ purpose\end{tabular}} &
  \begin{tabular}[c]{@{}c@{}}General-purpose language\\  understanding and generation\end{tabular} &
  2020 &
  No \\ \cline{1-3} \cline{5-7} 
GPT-4 \cite{GPT4MODEL} &
  Decoder-only transformer &
  \begin{tabular}[c]{@{}c@{}}Not publicly \\ disclosed\end{tabular} &
   &
  \begin{tabular}[c]{@{}c@{}}General-purpose image and language\\  understanding and text generation\end{tabular} &
  2023 &
  No \\ \cline{1-3} \cline{5-7} 
LLAMA \cite{LLAMAMODEL} &
  Decoder-only transformer &
  \begin{tabular}[c]{@{}c@{}}7B/13B/\\ 30B/65B\end{tabular} &
   &
  \begin{tabular}[c]{@{}c@{}}General-purpose language\\  understanding and generation\end{tabular} &
  2023 &
  Yes \\ \cline{1-3} \cline{5-7} 
PaLM \cite{PALMMODEL} &
  Decoder-only transformer &
  540B &
   &
  \begin{tabular}[c]{@{}c@{}}General-purpose language\\  understanding and generation\end{tabular} &
  2023 &
  \begin{tabular}[c]{@{}c@{}}Not officially \\ open-source\end{tabular} \\ \cline{1-3} \cline{5-7} 
BLOOM \cite{BLOOM} &
  Decoder-only transformer &
  176B &
   &
  \begin{tabular}[c]{@{}c@{}}General-purpose language\\  understanding and generation\end{tabular} &
  2022 &
  Yes \\ \hline
GATO \cite{GATOMODEL} &
  \begin{tabular}[c]{@{}c@{}}Decoder-only transformer\\ with ResNet patch\end{tabular} &
  1.17B &
  \multirow{5}{*}{Multi-modal} &
  \begin{tabular}[c]{@{}c@{}}Works as a multi-modal,\\  multi-task generalist agent\\ especially for robots\\ (images, text, joint torques,\\ button presses)\end{tabular} &
  2022 &
  Yes \\ \cline{1-3} \cline{5-7} 
ImageBind \cite{IMAGEBIND} &
  \begin{tabular}[c]{@{}c@{}}Modality-specific\\  encoders\end{tabular} &
  \begin{tabular}[c]{@{}c@{}}Not explicitly\\  disclosed\end{tabular} &
   &
  \begin{tabular}[c]{@{}c@{}}Combine data from six modalities, \\ including text, video, audio, depth, \\ thermal, and inertial measurement \\ unit (IMU), into a single embedding\\  space\end{tabular} &
  2023 &
  Yes \\ \cline{1-3} \cline{5-7} 
GPT-4o \cite{GPT4O} &
  Decoder-only transformer &
  \begin{tabular}[c]{@{}c@{}}Not explicitly\\  disclosed\end{tabular} &
   &
  \begin{tabular}[c]{@{}c@{}}General-purpose agent handling \\ multiple modalities including\\  text, audio, video, and images\end{tabular} &
  2024 &
  No \\ \cline{1-3} \cline{5-7} 
Gemini \cite{GEMINIMODEL} &
  \begin{tabular}[c]{@{}c@{}}Multimodal encoder\\ and transformer\end{tabular} &
  \begin{tabular}[c]{@{}c@{}}Varying sizes\\ for four \\ different\\ versions\end{tabular} &
   &
  \begin{tabular}[c]{@{}c@{}}General-purpose model handling\\ multiple modalities including\\ text, code, audio, image \\ and video\end{tabular} &
  2023 &
  No \\ \cline{1-3} \cline{5-7} 
DeepSeek R1 \cite{DEEPSEEK} &
  Decoder-only transformer &
  \begin{tabular}[c]{@{}c@{}}Not explicitly\\  specified, \\ around 671B\end{tabular} &
   &
  \begin{tabular}[c]{@{}c@{}}General-purpose agent handling \\ multiple modalities including\\  text, audio, and images\end{tabular} &
  2025 &
  Yes \\ \hline
\end{tabular}
}
\vspace{-10pt}
\end{table*}

\vspace{-5pt}
\subsection{Well-known Foundation Models and Wireless-Specific Foundation Models}

Throughout the evolution of FMs, several well-known models have emerged, marking significant milestones in the development of FMs. Table \ref{table-FMs} summarizes 17 well-known FMs, providing details about model architectures, sizes, functionalities, release years, and whether they are open source. These models cover diverse application domains, including language processing models, vision models, code generation models, general-purpose models, and multi-modal models. For multi-modal models, we only include FMs capable of processing data across more than three modalities, distinguishing them from vision-language models. Notably, some FMs may be applicable to multiple domains. For instance, some general-purpose FMs also function as language processing models, while some multi-modal models are also pre-trained as general-purpose FMs. 

As shown in the table, early FMs, such as BERT \cite{BERTMODEL} and T5 \cite{T5MODEL}, primarily focused on NLP tasks. Later, they were expanded to visual tasks. For visual tasks, early models like CLIP \cite{radford2021learning} and ViT \cite{ViTMODEL} were designed mainly for understanding visual information. However, with advancements in generative AI, later visual models, such as DALL-E \cite{DALLE2MODEL} and Stable Diffusion \cite{STABLEDIFFUSION}, enabled capabilities for visual generation tasks.

As FMs evolved, general-purpose models began to emerge, handling tasks in a language-based format (e.g., GPT-3 \cite{LanguageModelsAreFew-shotLearners}, LLaMA \cite{LLAMAMODEL}) or a combined visual and language format (e.g., GPT-4 \cite{GPT4MODEL}), commonly referred to as LLMs. These models are capable of performing a variety of tasks, including question-answering, problem-solving, control, and reasoning. Over time, FMs have been expanded towards more modalities, such as audio, video, and various sensor inputs.

Although FMs have been developed across various application domains, there are currently only a limited number of FMs specifically pre-trained for wireless communication. Large Wireless Model (LWM) proposed in \cite{LargeWirelessModel} is claimed by its developers as the first FM for wireless channels. LWM is based on transformer architecture and pre-trained in a self-supervised manner on large-scale wireless channel datasets. It excels in generating context-aware embeddings from raw wireless channel data and is applicable to downstream wireless network management tasks such as beam prediction and LOS classification. Both the model and its training data are open source. Another open-source mobile network-specialized LLM, Mobile-LLaMA, was developed as an instruction-fine-tuned variant of the LLaMA-2-13B model \cite{LLAMA2MODEL} and can perform three mobile network analysis functions: packet analysis, IP routing analysis, and performance analysis \cite{MobileLLAMA}.

In addition to LWM and Mobile-LLaMA, some recent research works have fine-tuned existing FMs or LLMs for text-based Telecom tasks. TeleRoBERTa is proposed \cite{TeleRoberta} as an adaptation of the RoBERTa base model in the Telecom domain. It is trained on a large corpus of text collected from in-domain sources such as 3GPP specification and can be used for Telecom-related question-answering tasks.  

\vspace{-10pt}
\subsection{Deployment of Foundation Models in Wireless Networks}

Given the large size of FMs, practical deployment becomes a prerequisite for fine-tuning and inference of FMs in wireless networks. The deployment method of FMs largely depends on the hardware support of network nodes, the complexity of the FMs and application requirements. In existing works, four deployment methods have been discussed, including cloud deployment, edge deployment, on-device deployment, and collaborative deployment: \begin{itemize}
    \item Cloud deployment: Cloud deployment is the most straightforward method for the deployment of FMs \cite{LargeLanguageModelForTelecommunications}. Cloud servers typically provide a range of configurations with varying hardware capabilities to support computationally intensive FMs. Therefore, cloud deployment is suitable for scenarios where high computational power is required. When deployed in a cloud environment, FM-based services can choose configurations and distribute model operations across multiple processing units with model parallelism to improve cost-efficiency \cite{AutomatingCloudDeployment}. However, cloud deployment presents several challenges, particularly in terms of latency and security. Information sharing with the cloud server raises privacy concerns. Moreover, communication with the cloud server introduces extra end-to-end latency, which makes cloud deployment unsuitable for latency-critical applications. 
    
    \item Edge deployment: Another deployment method is to deploy FMs at the edge of the network and integrate them with mobile edge computing nodes. This approach brings FMs closer to the data source, resulting in reduced latency and backhaul bandwidth usage compared to cloud deployment. It is particularly effective for edge-tailored models, which use techniques such as quantization, pruning, or distillation to reduce the computation cost. These optimizations allow for faster local inference, making edge deployment more efficient than relying on cloud-based execution. However, compared to the cloud server, mobile edge servers are typically resource-constrained, which makes edge deployment challenging. Various methods have been proposed to address this issue. For example, when multiple FMs with shared parameters or blocks are deployed at an edge node, the node can load a single copy of shared parameters, which can significantly reduce memory usage and minimize data swapping time \cite{GemelModelMerging}.
    
    \item On-device deployment: With the increasing capabilities of mobile devices, such as the integration of graphics processing units (GPUs), another deployment option has emerged as running FMs directly on devices. The on-device deployment method processes tasks locally, which can reduce the service latency. It also offers advantages, including high scalability and the ability to customize FMs based on specific user needs. However, the user devices face inherent limitations in computational and storage resources compared to cloud and edge deployment. Therefore, specialized techniques are essential to address these challenges, such as implementing cost-effective and fast inference using AI accelerators \cite{InferenceOptimization}, optimization for large models \cite{SpeedIsAllYouNeed} and hardware acceleration \cite{OnDeviceLanguageModels}.
    
    \item Collaborative deployment: Collaborative deployment refers to the joint utilization of cloud servers, edge servers, and devices for fine-tuning and inference of FMs. For instance, \cite{DeviceEdgeCooperativeFineTuning} presents a novel device-edge fine-tuning framework, enabling efficient and privacy-preserving fine-tuning services at the 6G network edge. Similarly, \cite{PushingLargeLanguageModels} proposes a device-server co-inference method and leverages the split inference technique to offload the computation from devices to a server via layer-wise model partitioning. 
    Such methods can simultaneously solve both the privacy concerns associated with cloud deployment and the computational limitations of on-device deployment. By integrating various network nodes, collaborative deployment balances the trade-offs between data privacy, computational resources, and real-time processing demands and offers more flexible FM-based applications to wireless networks. 
\end{itemize}

Efficient deployment of LWMs thus requires careful consideration of latency, privacy, resource availability, and application requirements. Table \ref{tab:compDeployment} summarizes these deployment paradigms, highlighting their strengths, limitations, and suitable application scenarios.

Beyond classification, the practical implementation of these models, especially in edge, on-device, and collaborative settings, relies on specialized serving systems and optimization frameworks. For on-device deployment, this involves leveraging model compression techniques (e.g., quantization, pruning) and lightweight inference engines such as TensorFlow Lite, Core ML, or ONNX Runtime that are optimized for mobile hardware \cite{OnDeviceLanguageModels}. For edge and collaborative deployment, the challenge lies in efficiently managing and orchestrating the models. This is often handled by edge computing frameworks that manage containerized AI workloads, and inference serving systems like NVIDIA Triton Inference Server or TorchServe, which can be deployed at the edge. Furthermore, collaborative systems for split computing \cite{DeviceEdgeCooperativeFineTuning} and federated fine-tuning \cite{DeviceEdgeCooperativeFineTuning} are critical enablers, providing the software architecture to partition the model and manage the interaction between the device and the server, thereby making complex FMs feasible on the wireless edge.

\begin{table*}[h!]
\caption{\textsc{Comparison of Deployment Paradigms for Serving LWMs}}
\renewcommand{\arraystretch}{1.6} 
\setlength{\tabcolsep}{6pt} 
\begin{tabular}{| >{\centering\arraybackslash\bfseries}m{3.0cm}
             | >{\centering\arraybackslash}m{4.0cm}
             | >{\centering\arraybackslash}m{4.5cm}
             | >{\centering\arraybackslash}m{5.5cm} |}
\hline
\textbf{Deployment Mode} & \textbf{Strengths} & \textbf{Weaknesses / Constraints} & \textbf{Suitable Application Scenarios} \\
\hline
Cloud deployment & Very large compute and memory available; easy to update and maintain; high model capacity \cite{LargeLanguageModelForTelecommunications}. & Higher latency; greater data transfer; weaker privacy guarantees; dependency on reliable network backhaul. & Tasks where latency is less critical and models need high capacity --- e.g. complex analytics, batch processing, or services aggregating data from many nodes \cite{AutomatingCloudDeployment}. \\
\hline
On-device deployment & Low latency; maximal data privacy; reduced dependence on network connectivity. & Limited compute/memory/energy; constraints on model size; harder to update or maintain large models. & Use cases requiring privacy or offline operation --- e.g. personal devices, IoT sensors with sensitive data, emergency or remote operations \cite{InferenceOptimization, SpeedIsAllYouNeed, OnDeviceLanguageModels}. \\
\hline
Edge deployment & Compromise between cloud and on-device: better compute than devices; lower latency than cloud; some privacy benefits. & Resource constraints (though better than on-device), management complexity; possibly heterogeneity of edge resources \cite{GemelModelMerging}. & Real-time or near real-time services in wireless networks like autonomous driving, AR/VR, networked control systems, and some smart city applications. \\
\hline
Collaborative / split / co-inference deployment & Allows partitioning model computation among device, edge, cloud; can optimize trade-offs of latency, privacy, energy, and resource usage \cite{DeviceEdgeCooperativeFineTuning, PushingLargeLanguageModels}. & Complexity in scheduling, partitioning, communication overhead; need to manage consistency, caching, straggler issues, etc. & Applications needing both strong privacy/security and low latency; large models that cannot fully reside on device; scenarios with intermittent connectivity or variable resource availability \cite{DeviceEdgeCooperativeFineTuning}. \\
\hline
\end{tabular}
\label{tab:compDeployment}
\end{table*}

\section{Foundation Model-Enabled Multi-Modal Contextual Information Understanding in Wireless Networks}
\label{s3}

In the context of wireless networks, the term multi-modal refers to the integration and joint analysis of heterogeneous data sources, such as visual, spatial, and CSI, rather than the isolated treatment of each modality. Multi-modal fusion enables networks to capture richer contextual information by combining complementary perspectives from different modalities. Given this, one primary application of FMs in this domain is the interpretation of this multi-modal contextual information. Wireless networks can be considered sensor networks if network nodes function as sensors. By analyzing this sensor data, changes in the propagation environment caused by moving objects can be captured and used to optimize wireless network management \cite{MultimodalWirelessNetworksCommunication}. For example, recent studies have demonstrated how visual data and wireless communication features can be fused to improve proactive beamforming decisions in vehicular networks \cite{zhang2024integrated}. Similarly, multi-modal representation learning techniques such as vector-quantized variational autoencoders (VQ-VAE) have been employed to fuse diverse data modalities—including images, spectrograms, and wireless CSI—for efficient transmission and robust performance in communication \cite{bocus2023streamlining}. These works highlight that the strength of multi-modal FMs lies in their ability to learn shared representations across modalities, thereby enabling more adaptive and context-aware wireless network management. This section examines how FMs can be applied to analyze and generate embeddings for these various types of contextual data, including visual data, graph data, point cloud data, and other modalities.

\subsection{Vision-based Wireless Applications Enhanced by Foundation Models} \

Vision data, typically referring to RGB images captured by visual sensors like cameras, plays a crucial role in wireless communication systems. Many devices in wireless networks, such as 5G smartphones, self-driving cars, and virtual/augmented reality headsets, are equipped with cameras to support advanced functionality. Additionally, researchers have proposed incorporating cameras at millimeter wave (mmWave) BSs, highlighting potential gains in obtaining sensing information and advanced network management \cite{jiang2022camera,MillimeterWaveBaseStations, zhang2022camera}. Therefore, visual data is accessible in wireless networks and should be utilized to overcome non-trivial wireless communication challenges.

\begin{figure}[!t] \centering \includegraphics[width=3.5in]{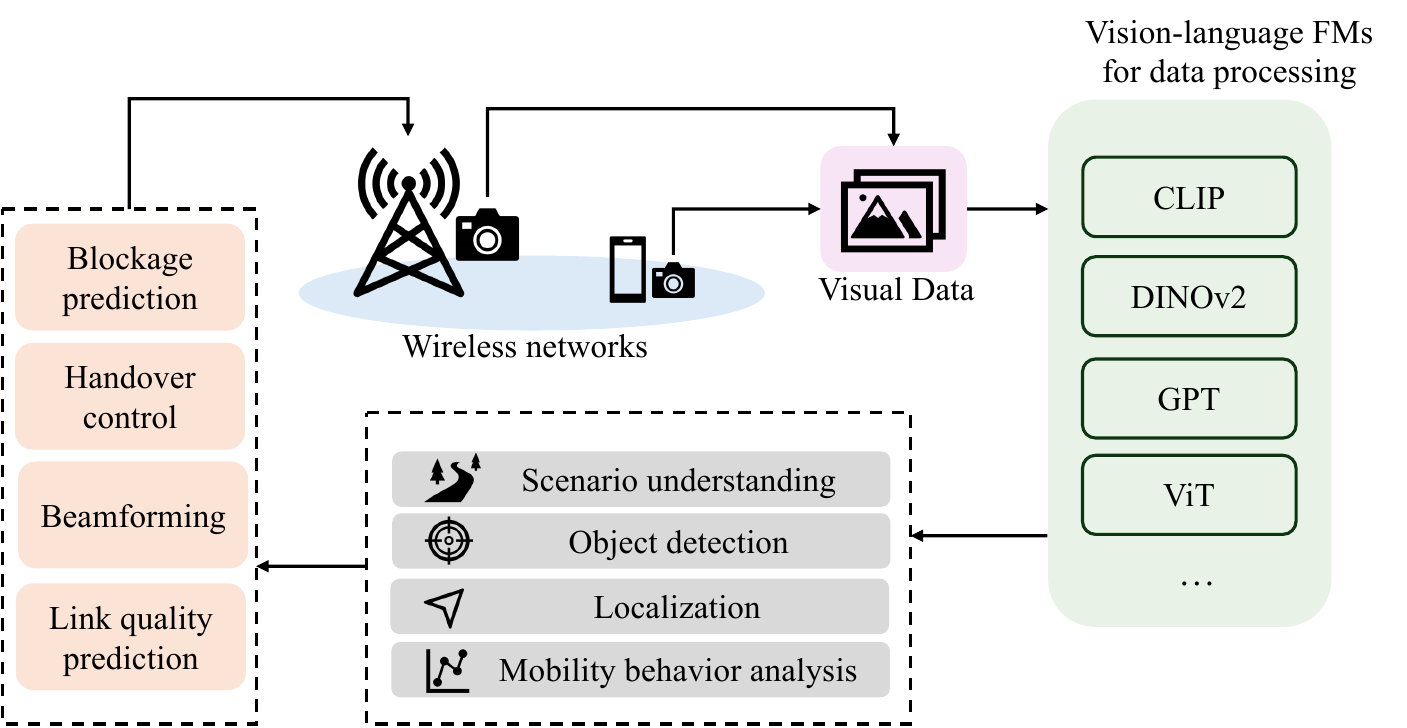} \caption{FM-enhanced visual data analysis and applications in wireless networks} \label{Visual} \vspace{-10pt} \end{figure}

In Fig. \ref{Visual}, the FM-enhanced visual data analysis and applications in wireless networks are presented. This subsection explores the applications from two perspectives. First, we examine the capabilities of FMs, exploring what types of visual information in wireless networks can be analyzed by FMs. Second, we highlight possible vision-aided wireless applications and discuss how they can be enhanced by integrating the contextual visual information extracted by FMs.

\subsubsection{Foundation Model Capabilities for Visual Information Extraction} 

We categorize FM capabilities for wireless visual data analysis into three primary functions: semantic scene understanding, object detection and localization, and temporal motion analysis.

\begin{itemize} 
\item \textbf{Semantic Scene Understanding:} The primary capability of visual FMs is to perform label-free scenario understanding and extract high-level semantic information from complex visual scenes. Vision-language models such as CLIP \cite{radford2021learning} have demonstrated exceptional zero-shot transfer capabilities by learning joint embeddings of visual and textual representations. The model achieves strong zero-shot classification performance across over 30 different computer vision datasets, matching the accuracy of a ResNet-50 trained on ImageNet without using any of the 1.28 million ImageNet training examples \cite{radford2021learning}. Recent work leveraging CLIP for open-world object detection has shown significant performance improvements on challenging benchmarks such as LVIS.Self-supervised ViTs such as DINOv2 \cite{oquab2023dinov2} provide another powerful approach for semantic understanding without requiring labeled data. DINOv2, trained on 142 million curated images, achieves 82.0\% k-NN evaluation accuracy and 84.5\% linear evaluation accuracy on ImageNet \cite{oquab2023dinov2}. The model demonstrates strong transferable visual features that work across various computer vision tasks without fine-tuning, including image classification, semantic segmentation, and depth estimation. These capabilities are particularly valuable for wireless scenarios where labeled data is scarce.

\item \textbf{Object Detection and Localization:} FMs enable precise identification and localization of mobile devices and network components in visual data. The Segment Anything Model (SAM) \cite{kirillov2023segment} provides promptable segmentation capabilities that can be adapted for wireless device detection. SAM was trained on over 1 billion masks from 11 million images and demonstrates impressive zero-shot segmentation performance across diverse image distributions. Grounding DINO \cite{liu2023grounding} combines transformer-based detection with language grounding, achieving 52.5 AP on COCO zero-shot detection benchmark and 63.0 AP with COCO fine-tuning, enabling text-prompt-based identification of wireless devices without task-specific training. For wireless-specific applications, vision-aided positioning demonstrates that FM-aided object detection can identify device locations in images for positioning systems. Experimental work by \cite{alrabeiah2020deep} shows that vision-based approaches can detect and localize transmitters in indoor scenarios for beam prediction tasks. This capability is especially important for devices lacking access to GPS in indoor environments.

\item \textbf{Temporal Motion Analysis:} FMs can analyze device mobility patterns through temporal visual data processing. Pre-trained video understanding models enable trajectory prediction and motion forecasting for wireless applications. The work in \cite{jiang2022computer} demonstrates encoder-decoder architectures for vision-aided beam tracking that exploit temporal information from image sequences to predict future beams in mmWave systems. By utilizing time-continuous image data for object detection, wireless operators can track device mobility, reconstruct past trajectories, and anticipate future locations for proactive resource planning.
\end{itemize} 

\subsubsection{FM-Enhanced Wireless Applications} Based on the FM capabilities identified above, we systematically categorize wireless applications into four domains: link reliability enhancement, mobility management, beam management, and quality prediction.

\begin{itemize}
    \item Blockage prediction: Terahertz networks in the B5G era mainly rely on LOS links and are highly susceptible to sudden blockages. Therefore, performing accurate blockage prediction can enhance the reliability of the communication \cite{VisionAided6GWireless}. With advanced scenario understanding capabilities, pre-trained FMs can be utilized to extract critical embeddings from images to aid in blockage prediction or even directly perform the prediction.
    
    \item Handover control: Visual information is critical for effective mobility management and handover control in wireless networks \cite{SensingAndComputerVision}. FMs can process visual sensing information to extract and analyze the mobility patterns and geometric information of the target devices. These insights provide valuable contextual data for cell association prediction and can be leveraged for handover control.
    
    \item Beamforming: Compared with traditional RF signal-based beam management, vision-aided beam management offers many advantages, including enhanced beamforming gain and reduced latency. This improvement is that the beam direction can be directly set without relying on codebook quantization and feedback delays \cite{TowardIntelligentMillimeter}. In vision-aided beam management tasks, FMs can function as object detectors, identifying the location of mobile devices and thereby assisting in determining the optimal beam transmission direction.
    
    \item Link quality prediction: The quality of wireless links is strongly influenced by changes in the surrounding environment, particularly when the communication frequencies shift to higher bands. FMs can be used to enhance the ability of wireless devices to perceive their own position and the surrounding environment, and this information can be leveraged to predict the potential impact of nearby mobile objects on communication channels. Therefore, the vision-based perception information can be effectively utilized to predict wireless link quality \cite{UsingVisionBasedObjectDetection}.
\end{itemize}

\vspace{10pt}

\subsection{Graph-based Wireless Network Analysis Enhanced by Foundation Models}

Graph representations serve as a simple framework for modeling wireless networks, as they naturally capture the inherent topology of these networks. In recent years, graph-based analysis has become increasingly important for optimizing and managing wireless network operations \cite{shen2022graph,GraphNeuralNetworksForWirelessNet}. Graphs consist of two fundamental components: nodes and edges. In the context of wireless networks, nodes typically correspond to network elements such as BSs, user equipment, and Internet of Things (IoT) devices and the edge of the graph represent the communication links that connect these nodes. These edges are weighted by metrics such as CSI, received signal strength (RSS), signal-to-noise ratio (SNR), or other relevant factors and represent the interactions between network elements  \cite{GraphRepresentationLearningForWireless}. By capturing these relationships, graph representations offer a robust foundation for optimizing and effectively managing the wireless communication system \cite{AnOverviewOnTheApplication, EdgeGraphIntelligenceReciprocally}. 

Graph Foundation Models (Graph FMs) are designed to learn transferable knowledge from large-scale graph datasets, enabling generalization across diverse wireless scenarios with minimal fine-tuning. Their capabilities are built on advanced pre-training strategies, such as masked autoencoding and multi-task learning frameworks. These models address key challenges like scalability, using specialized architectures to reduce computational costs, and relational complexity, by using specific architectures like Heterogeneous Graph Transformers (HGT) \cite{hu2020heterogeneous} to manage multiple node and edge types. HGT was evaluated on the Open Academic Graph containing 179 million nodes and 2 billion edges, consistently outperforming state-of-the-art GNN baselines by 9\%-21\% across various downstream tasks. Furthermore, graph prompting techniques allow for rapid adaptation without full retraining, significantly reducing training overhead. In wireless networks, these capabilities are applied to solve optimization and management tasks.

Considering the capabilities of graph FMs and the information they can extract from wireless graph data, the potential applications of graph FM in wireless networks are detailed below:

\subsubsection{Fundamental AI works about graph FMs}

Graph FMs have emerged as a significant advancement in the graph domain. They are designed to learn transferable knowledge from large-scale graph datasets that can be applied across different wireless network scenarios \cite{TowardsGraphFoundationModels, PositionGraphFoundationModels}. This subsection examines how graph FMs can be applied to analyze network topologies by exploring the fundamental aspects of graph construction and discussing the applications of graph FMs in wireless networks.

The inherent invariance properties of graphs are especially important in wireless networks since the performance of wireless networks is heavily influenced by the connectivity patterns and relative positioning of devices. \cite{EdgeFMLeveraging} introduces EdgeFM, an FM designed for open-set learning on the edge of the graphs, and it demonstrated that graph FMs could effectively learn invariant information in graphs. Similarly, the research in \cite{TowardsAFoundationModel} applies an FM to time series data, leveraging graph-based representations of nodes and edges to capture both local and global network characteristics. These studies underscore the critical role of FMs in extracting graph information and advancing the analysis and prediction of network behaviors and performance.

GraphMAE \cite{hou2022graphmae} presents a masked graph autoencoder for self-supervised graph learning that addresses key limitations of traditional graph autoencoders. Instead of reconstructing graph structures, GraphMAE focuses on feature reconstruction with a masking strategy and scaled cosine error metric. The model demonstrates strong performance across 21 datasets on node classification, graph classification, and molecular property prediction tasks, achieving results competitive with or exceeding contrastive learning methods while maintaining a simpler training paradigm. This generative approach offers an effective alternative to complex data augmentation strategies while learning robust graph representations.

In addition, recent research has emphasized the importance of scalable frameworks and benchmarks to evaluate the performance and efficiency of graph FMs across various tasks, such as node classification, link prediction, and node clustering. \cite{GraphFMAComprehensive} introduces GraphFM as a comprehensive benchmark designed to assess the capabilities of graph FMs across diverse datasets and graph-related tasks. Building on this, \cite{GraphFMAScalable} proposes a scalable framework for multi-graph pre-training that employs a perceiver-based encoder with learned latent tokens. This approach compresses domain-specific features into a shared latent space for multi-graph multi-task pre-training, showing competitive performance across diverse graph datasets across various graph types and tasks. These graph FMs with strong capabilities can be applied to tasks related to graph-based wireless networks.

\subsubsection{Applications of graph FMs in wireless networks}
Considering the capabilities of graph FMs and the information they can extract from wireless graph data, the potential applications of graph FM in wireless networks are detailed below:

\begin{itemize} 

    \item \textbf{Mobility Management and Handover Optimization:} Modeling the dynamic topology of UEs and BSs as a graph enables precise mobility management. GNN-based approaches have demonstrated strong performance in predicting handover events and cell associations. Spatio-temporal graph networks are used to predict cell association and UE trajectories, achieving high accuracy in predicting future network states. These models can predict cell associations several seconds ahead and track vehicular mobility with low displacement error, leading to significant reductions in unnecessary handovers through intelligent differentiation between temporary signal fluctuations and persistent mobility patterns.

    \item \textbf{Interference Management and Resource Allocation:} By representing interference patterns as a graph, GNN-based approaches achieve significant gains in wireless resource allocation. \cite{shen2020graph} demonstrates that GNNs enable scalable power control in k-user interference channels, providing solutions that generalize across different network sizes without retraining. The approach leverages the permutation equivariance property of GNNs to handle networks of varying topologies. Recent work by \cite{zhang2021scalable} shows that GNN-based power control in heterogeneous wireless networks achieves performance within 1\% of optimal solutions while providing 1000× speedup over iterative optimization methods. Furthermore, \cite{lo2022egraphsage} introduces E-GraphSAGE for IoT network intrusion detection, achieving F1-scores of 1.0 on NF-ToN-IoT and 0.97 on NF-BoT-IoT datasets by incorporating both flow-level features and network topology patterns. These approaches leverage the graph structure to model complex interference relationships and optimize resource allocation decisions.

    \item \textbf{Anomaly Detection and Network Security:} Graph FMs excel at identifying anomalous patterns indicating malicious activities or network faults. \cite{protogerou2021graph} proposes a distributed anomaly detection scheme using multi-agent systems where each agent implements a GNN to detect DDoS attacks in IoT networks. The approach models network flow-level information as graph structures and demonstrates effectiveness in detecting propagating cyber-attacks across distributed fog nodes. \cite{lo2022egraphsage} achieves binary classification F1-scores approaching 1.0 on multiple IoT security datasets (ToN-IoT, BoT-IoT) for detecting various attack types including DDoS, reconnaissance, and data theft. Recent work by \cite{ahmad2025graphfedai} introduces GraphFedAI, a federated learning framework integrating adaptive session-based graph modeling with GNNs for DDoS detection in IoT environments, achieving robust performance while preserving privacy across heterogeneous networks. Federated graph learning approaches have been proposed to address privacy concerns while maintaining high anomaly detection accuracy.
\end{itemize}

\vspace{10pt}

\subsection{3D Point Cloud Information Analysis in Wireless Networks}

In wireless networks, 3D point cloud data are usually gathered through advanced technologies such as LiDAR and radar. These data can provide rich spatial details and enable networks to understand the geometric and structural characteristics of the contextual environment. Such information enhances the performance of wireless communications by facilitating precise modeling, improved adaptability, and optimized connectivity in complex wireless environments. It is particularly valuable in scenarios such as vehicular networks and urban connectivity, where accurate environmental modeling is necessary \cite{DeepLearningFor3D}.

FMs can utilize 3D point clouds for tasks like segmentation, classification, and scenario understanding, or refine these spatial representations to better suit the needs of specific applications in wireless networks. For instance, \cite{PointSam} extends the SAM \cite{kirillov2023segment} to the 3D domain and proposed the Point-SAM framework, which utilizes a transformer-based architecture tailored for point clouds. Its ability to operate with minimal task-specific training highlights its adaptability in real-world applications. By enabling flexible segmentation of 3D point clouds, Point-SAM can be used to enhance diverse applications in wireless communication scenarios, such as distinguishing between static and dynamic objects. Another FM for 3D point clouds, PointLLM, extended LLMs to interpret spatial data, bridging the gap between language-based reasoning and point cloud analysis \cite{xu2024pointllm}. In \cite{PointBert}, a novel approach called Point-BERT was introduced to pre-train transformers for 3D point cloud data. The framework employed a masked point modeling strategy, where parts of the point cloud are masked, and the model learns to reconstruct the missing regions. 

On the other hand, 3D point cloud analysis has found diverse applications in wireless networks that can enhance performance and adaptability across various scenarios. For instance, the development of spatio-temporal 3D point clouds from WiFi CSI data enables the replication of physical environments, facilitating advanced applications in smart cities and healthcare monitoring \cite{SpatioTemporal3D}. Other 3D point cloud-assisted applications in wireless networks include mmWave communication-assisted localization \cite{MillimeterWaveWirelessCommunicationAssisted}, radio propagation modeling \cite{RayProNet} and resource allocation for wireless metaverse services \cite{QoEAnalysisAndResource}.

Recent research has begun to specifically combine FMs with 3D point cloud data for wireless network tasks. 
LiDAR-LLM leverages LLMs to interpret raw LiDAR data, enabling a comprehensive understanding of outdoor 3D scenes, which can be applied to tasks such as 3D captioning and question answering \cite{LiDARLLM}. Additionally, the NetOrchLLM framework aims to optimize wireless network orchestration with LLMs, considering inputs from various sensors such as RADAR, LiDAR, and cameras \cite{NetOrchLLM}.
\cite{jiang2022lidar} presents the first large-scale real-world evaluation of LiDAR-aided beam prediction in mmWave vehicle-to-infrastructure (V2I) communications using the DeepSense 6G dataset \cite{alkhateeb2023deepsense}. The ML model leveraging LiDAR sensory data achieves 95\% beam prediction accuracy with over 90\% reduction in beam training overhead compared to exhaustive search. The LiDAR-aided beam tracking achieves comparable accuracy to a baseline with perfect knowledge of previous optimal beams, without requiring any prior beam information or calibration. This demonstrates the practical value of leveraging 3D point cloud data for wireless beam management in highly-mobile scenarios.

Collectively, these advances demonstrate the significant and measurable potential of leveraging FMs to unlock novel capabilities from 3D point cloud data in wireless communication systems, paving the way for more intelligent, adaptive, and spatially aware networks.

\subsection{Other Modalities and Data Sources in Wireless Networks}

In addition to visual, graph-based, and 3D point cloud data, wireless networks leverage a variety of other modalities to enhance their performance and adaptability. These modalities include Integrated Sensing and Communication (ISAC) data, RF data, and network information. By leveraging their ability to process and unify diverse data types, FMs can be employed to analyze and optimize these modalities, contributing to improved accuracy, efficiency, and decision-making in wireless systems. The details are given as follows:

\subsubsection{ISAC data}

ISAC is a technique that integrates the dual functionalities of wireless communication and environmental sensing. FMs trained on ISAC data can predict environmental changes, optimize spectrum usage, and enable real-time adjustments in dynamic and complex environments \cite{LargeLangugaeModelBasedMultiObjective}. \cite{LargeLanguageModelsEmpowerMultimodalIntegrated} highlights the potential of FMs in enhancing ISAC within wireless networks. By effectively processing and fusing multi-modal data, including sensing and communication signals, FMs optimize resource allocation and improve situational awareness. The integration of FMs in wireless environments improves the efficiency, scalability, and reliability of ISAC frameworks in next-generation networks.

\subsubsection{RF data}
RF signals serve dual purposes in wireless systems: enabling location-based services (RF localization) and supporting environmental perception (RF sensing). In localization, FM approaches for RF positioning have demonstrated significant improvements. \cite{ott2024radio} proposes the Radio FM, a self-supervised learning framework that pre-trains a transformer neural network on 5G channel measurements. The model learns spatiotemporal patterns from unlabeled channel impulse response (CIR) data without requiring reference positions. When fine-tuned for localization tasks, the Radio FM achieves state-of-the-art accuracy with a mean localization error of 0.8 meters, representing approximately 45\% improvement over non-pre-trained DL baselines. Critically, the model requires 10 times less labeled reference data and significantly reduces the time from training to deployment compared to traditional fingerprinting methods, demonstrating the practical value of FM approaches for RF localization in 5G indoor environments.

For RF sensing, generative AI models interpret signal perturbations caused by human activity or environmental changes, achieving high accuracy in activity recognition tasks and enabling synthesis of realistic RF data to improve classifier performance in data-scarce settings. Additionally, FMs are advancing RF fingerprinting for security applications, with transformer-based models for device fingerprinting attaining over 99\% accuracy in device identification by learning unique RF emission signatures.

\subsubsection{Network usage data}
Wireless networks continuously generate rich usage data, including traffic patterns, resource utilization metrics, and device interaction logs. FMs are being developed to analyze this complex, high-dimensional data for real-time network optimization. General-purpose network FMs are pre-trained on massive corpora of network documentation, logs, and telemetry to function as autonomous network administrators. For domain-specific tasks, specialized models fine-tuned on wireless network data have achieved high accuracy in root-cause analysis tasks, significantly outperforming general-purpose models. By capturing intricate dependencies among network parameters, such models enable more accurate predictions, proactive bottleneck detection, and dynamic configuration adjustments, significantly improving QoS and operational efficiency.

\begin{table*}[]
\centering
\renewcommand\arraystretch{1.2}
\caption{A Summary of Existing FMs for Time-Series Prediction or Frameworks for Developing FMs for Time-Series Prediction}
\label{tab:time}
\resizebox{2\columnwidth}{!}{%
\begin{tabular}{|c|c|c|c|c|c|}
\hline
\textbf{Model Name}                        & \textbf{Model Input}                                                                                                    & \textbf{Main Architecture}                                                                                                            & \textbf{Training Method}                                                                                                                 & \textbf{Data Source}                                                                                                                     & \textbf{Capabilities}                                                                                                                                                             \\ \hline
TimeGPT-1 \cite{garza2023timegpt}                  & \begin{tabular}[c]{@{}c@{}}Historical values of the target \\ values and additional \\ exogenous variables\end{tabular} & \begin{tabular}[c]{@{}c@{}}Encoder-decoder \\ transformer-based \\ architecture\end{tabular}                                          & Trained from scratch                                                                                                                     & \begin{tabular}[c]{@{}c@{}}A collection of publicly \\ available time series from \\ a broad array of domains\end{tabular}               & \begin{tabular}[c]{@{}c@{}}Generate accurate \\ predictions for \\ diverse datasets\end{tabular}                                                                                  \\ \hline
TimeDiT \cite{TimeDIT}                     & \begin{tabular}[c]{@{}c@{}}Multivariate multi-resolution \\ time-series sequences \\ with missing values\end{tabular}   & Diffusion transformer                                                                                                                 & \begin{tabular}[c]{@{}c@{}}Masked self-supervised \\ pre-trained from scratch\end{tabular}                                               & Chronos dataset                                                                                                                          & \begin{tabular}[c]{@{}c@{}}Various time-series tasks, \\ including forecasting, \\ imputation, anomaly \\ detection, and synthetic\\  data generation.\end{tabular}               \\ \hline
Lag-Llama \cite{lagllama}                  & Univariate time series                                                                                                  & LlaMA architecture \cite{LLAMAMODEL}                                                                                                  & Trained from scratch                                                                                                                     & \begin{tabular}[c]{@{}c@{}}A collection of time series\\  from the Monash Time \\ Series Repository \cite{MonashTimeSeries}\end{tabular} & \begin{tabular}[c]{@{}c@{}}Univariate time \\ series forecasting\end{tabular}                                                                                                     \\ \hline
TimesFM \cite{ADecoderOnlyFoundationModel} & Univariate time series                                                                                                  & Decoder-only transformer                                                                                                              & Trained from scratch                                                                                                                     & \begin{tabular}[c]{@{}c@{}}Real-world data\\ and synthetic data\end{tabular}                                                             & \begin{tabular}[c]{@{}c@{}}Univariate time \\ series forecasting\end{tabular}                                                                                                     \\ \hline
MOMENT \cite{goswami2024moment}                & \begin{tabular}[c]{@{}c@{}}Univariate time series of \\ a fixed length\end{tabular}                                     & Encoder-only transformer                                                                                                              & \begin{tabular}[c]{@{}c@{}}Trained from scratch \\ using masked\\  time-series modelling\end{tabular}                                    & \begin{tabular}[c]{@{}c@{}}Time-series data \\ from diverse domains\end{tabular}                                                         & \begin{tabular}[c]{@{}c@{}}General-purpose time \\ series analysis, including\\  forecasting, imputation, \\ anomaly detection,\\  and synthetic \\ data generation.\end{tabular} \\ \hline
LLM4TS \cite{chang2024llm4ts}                       & Multivariate time series                                                                                                & GPT-2 architecture \cite{GPT2MODEL}                                                                                                   & \begin{tabular}[c]{@{}c@{}}Fine-tuning on pre-trained\\ GPT-2 \cite{GPT2MODEL}\end{tabular}                                              & \begin{tabular}[c]{@{}c@{}}Task-specific time series data,\\ such as Electricity and Weather\\ datasets\end{tabular}                     & \begin{tabular}[c]{@{}c@{}}Multivariate time \\ series forecasting\end{tabular}                                                                                                   \\ \hline
One fits all \cite{zhou2023onefitsall}             & Univariate time series                                                                                                  & \begin{tabular}[c]{@{}c@{}}Chosen FM backbone \\ with input embedding\end{tabular}                                                    & \begin{tabular}[c]{@{}c@{}}Fine-tuning on pre-trained \\ FMs such as \\ GPT-2 \cite{GPT2MODEL} and \\ BERT \cite{BERTMODEL}\end{tabular} & \begin{tabular}[c]{@{}c@{}}Task-specific time series data,\\ such as Electricity and Weather\\ datasets\end{tabular}                     & \begin{tabular}[c]{@{}c@{}}General-purpose time\\  series analysis, including\\  forecasting, imputation, \\ anomaly detection,\\  and synthetic data\\  generation.\end{tabular} \\ \hline
TEMPO \cite{cao2023tempo}                         & \begin{tabular}[c]{@{}c@{}}Multivariate numerical time \\ series with textual information\end{tabular}                  & GPT-2 architecture \cite{GPT2MODEL}                                                                                                   & \begin{tabular}[c]{@{}c@{}}Prompt tuning and low-rank \\ adaption (LORA) based on\\ pre-trained GPT-2\end{tabular}                       & \begin{tabular}[c]{@{}c@{}}Task-specific time series data,\\ such as Electricity and Weather\\ datasets\end{tabular}                     & \begin{tabular}[c]{@{}c@{}}Time-series representation\\  learning and forecasting\end{tabular}                                                                                    \\ \hline
GPT4MTS \cite{GPT4MTS}                     & \begin{tabular}[c]{@{}c@{}}Multivariate numerical time \\ series with textual information\end{tabular}                  & \begin{tabular}[c]{@{}c@{}}GPT-2 architecture \cite{GPT2MODEL}\\ with separate embedding \\ layers for two \\ modalities\end{tabular} & \begin{tabular}[c]{@{}c@{}}Fine-tune the positional \\ embeddings and \\ normalization\\  layers of GPT-2\end{tabular}                   & \begin{tabular}[c]{@{}c@{}}Derived from GDELT\\ dataset\end{tabular}                                                                     & \begin{tabular}[c]{@{}c@{}}Multivariate time \\ series forecasting\end{tabular}                                                                                                   \\ \hline
\end{tabular}
}
\vspace{-10pt}
\end{table*}

\vspace{-5pt}

\section{Foundation Models for Prediction Tasks in Wireless Networks}
\label{s4}

In this section, we explore how to leverage FMs to perform prediction tasks in wireless networks. We begin with a brief overview of various prediction tasks in wireless networks and the existing FMs developed specifically for prediction. Next, we discuss the application of FMs for traffic prediction and other prediction tasks within wireless network environments.

\vspace{-10pt}
\subsection{Foundation Models for Prediction}

Prediction tasks involve forecasting the value of a response variable that remains unobserved in future data. A critical focus of these tasks lies in how to effectively handle the time-series data. There are different types of time series in wireless networks, including standard time series and spatial time series. Standard time series consists solely of data in the time domain, whereas spatial time series incorporates both temporal data and the physical or logical topology of cellular networks in the spatial domain. In addition, according to the data type of the variable to be predicted, prediction tasks can be categorized into two types. The first type involves predicting continuous variables, such as wireless traffic prediction. The second type involves predicting discrete variables, such as blockage prediction, and this type is also regarded as time-series classification problems.

While traditional statistical methods and specialized DL architectures like GNNs or Transformers are effective, they are typically trained from scratch for a single, specific prediction task. This limits their flexibility. The fundamental nature of FMs offers a paradigm shift: pre-training on massive, diverse datasets to learn universal time-series representations. This pre-training captures generalized knowledge, enabling a single model to demonstrate general applicability across various tasks, often with minimal adaptation, such as in a zero-shot setting. This adaptability extends to handling varying historical data lengths, prediction horizons, and time granularities, pointing toward the potential for a single, unified prediction model.

Table \ref{tab:time} summarizes prominent FMs developed for time-series prediction \cite{garza2023timegpt, liu2023grounding, rasul2024lagllama, ADecoderOnlyFoundationModel, goswami2024moment} and representative frameworks for adapting LLMs to this domain \cite{chang2024llm4ts, zhou2023onefitsall, cao2023tempo,niu2023understanding}. 
These models highlight the trend toward generality, though their current scopes vary. Some models are trained from scratch to be general-purpose time-series analyzers, capable of forecasting, imputation, and anomaly detection, such as MOMENT \cite{goswami2024moment}. Others, like TimeGPT-1 \cite{garza2023timegpt} or Lag-Llama \cite{rasul2024lagllama}, are more focused on zero-shot forecasting. TimeGPT-1, trained on over 100 billion data points, demonstrates strong zero-shot transfer capabilities across diverse domains. Lag-Llama, a decoder-only transformer pretrained on approximately 8,000 univariate time series (352 million tokens) from 27 datasets across six domains, achieves strong zero-shot generalization and state-of-the-art performance when fine-tuned on downstream tasks. A significant trend is the adaptation of pre-trained LLMs, which inherently treat language as a time series. Frameworks like LLM4TS \cite{chang2024llm4ts} and TEMPO \cite{cao2023tempo} repurpose models like GPT-2 for forecasting. This approach cleverly leverages the FMs' existing strengths in handling missing data, integrating textual information, and providing interpretability, which are highly relevant for complex time-series tasks.

Despite this progress, a key challenge remains in applying general-domain FMs directly to wireless networks. Many models in Table \ref{tab:time} are pre-trained on generic time-series data and primarily accept numerical sequences as input. However, wireless prediction tasks are rarely unimodal; they require analyzing temporal patterns in conjunction with rich contextual information, such as user mobility, network topology, or signal quality indicators. This context is critical for accurate forecasting. Rather than being a fundamental limitation, this highlights the need for effective adaptation strategies. A promising solution is to leverage the multimodal capabilities of LLM-based FMs. Contextual information can be represented as natural language prompts or tokens, aligned with the numerical time-series data through mechanisms like linear projection and multi-head attention. This allows the FM to fuse and reason over both temporal and contextual data.

In the following subsection, we investigate the detailed application examples of FMs in traffic prediction tasks and other prediction tasks in wireless networks.

\subsection{Foundation Models for Wireless Traffic Analysis and Prediction}

Wireless traffic prediction is a fundamental prediction task in wireless networks. The growing number of wireless users and the variety of services produce a large amount of traffic, making it difficult to manage network resources. By accurately predicting traffic patterns, network operators can optimize resource allocation and scheduling strategies, contributing to the development of more intelligent and efficient wireless networks. However, this task is inherently difficult due to the complex patterns hidden in historical traffic data. These complexities highlight the need to use FMs to improve the accuracy and robustness of wireless traffic prediction.

There are three ways to perform wireless traffic prediction with FMs. The first approach is to train an FM specifically for wireless traffic prediction from scratch and then fine-tune it for specific tasks. The second approach is to leverage open-source pre-trained FMs and fine-tune them for wireless traffic prediction tasks. The third approach is to use pre-trained FMs directly for wireless traffic prediction without performing any fine-tuning.

\subsubsection{Train FMs from scratch} 
For the first approach, training an FM specifically for wireless traffic prediction from scratch involves two key steps: traffic tokenization and pre-training \cite{LensAFoundation}. The tokenization step plays a crucial role in handling the diversity of traffic data from various periods and diverse urban regions and converting them into a standardized format \cite{FoMoAFoundationModel}. The pre-training step is to enhance the understanding of FMs in wireless traffic data and make FMs learn rich and transferable representations for prediction from large-scale, heterogeneous traffic datasets. In the literature, some works have adopted this approach. For instance, FoMo is an FM designed for mobile traffic forecasting \cite{FoMoAFoundationModel}. It integrates the structure of diffusion models and transformers while employing a masked diffusion model with a self-supervised training strategy for pre-training. This model was then applied to optimize multiple aspects of network performance, including coverage, throughput, and energy consumption. Similarly, OpenCity is an open spatio-temporal FM for traffic prediction pre-trained on large-scale, heterogeneous traffic datasets \cite{OpenCityOpenSpatio}. Although OpenCity was designed for road traffic prediction rather than mobile traffic prediction, these two types of tasks share some similarities, allowing their model design methods to be applied interchangeably. 

In addition, some recent research has developed ML models for wireless traffic prediction based on transformer architecture. For instance, \cite{TransformerBasedWireless} has introduced a transformer-based method for predicting wireless network traffic in concise temporal intervals. Similarly, \cite{ASpatialTemporalTransformer} has proposed a transformer-based network that can excavate characteristics of cellular traffic for accurate cellular traffic prediction. These transformer-based models also have the potential to function as FMs.

Training traffic prediction FMs from scratch outperforms general-purpose time series models, such as TimeGPT-1 \cite{FoMoAFoundationModel}. However, the primary challenges include the limited availability of large-scale datasets and the significant computational costs.

\subsubsection{Fine-tune open-source pre-trained FMs}
The second approach, i.e. fine-tuning open-source pre-trained FMs for traffic prediction tasks, can make up for some of the shortcomings of the first approach. An example is seen in \cite{DeepTransferLearning}, where a model was first pre-trained by abundant data of a source city to obtain prior knowledge of mobile traffic dynamics. It was then fine-tuned to the prediction tasks in other target cities through the GAN-based cross-domain adapter. Additionally, several other studies have fine-tuned LLMs for traffic flow prediction tasks \cite{ExplainableTrafficFlowPrediction}. The fine-tuning process can enhance the capability of FMs to understand contextual information and data patterns in the new testing scenario, even with limited scenario-specific data.

\subsubsection{Leverage pre-trained FMs directly}
In the third approach, some LLMs are applied directly to wireless traffic prediction tasks without any fine-tuning. One important reason is that the input data for wireless traffic prediction can be structured in a format similar to natural language, making it feasible to leverage pre-trained LLMs for this task. Many pre-trained LLMs are equipped with mathematical problem-solving capabilities, which enable them to analyze complex patterns in traffic data. These strengths make LLMs a potential approach for wireless traffic prediction. For instance, in \cite{SelfRefinedGenerative}, TrafficLLM is designed to leverage pre-trained LLMs for wireless traffic prediction using ICL and self-refinement, without the need for additional parameter fine-tuning or model training. ICL is a critical method of guiding LLM responses by including a few known examples or demonstrations in the prompts. By leveraging ICL, LLMs can be directly used for specific tasks without the need for fine-tuning.

Choosing from the three above-mentioned approaches to perform traffic prediction tasks with FMs depends on the availability of suitable FMs, data, and computational resources. Given the success of FM-based wireless traffic prediction demonstrated in the aforementioned literature, the next subsection will explore the potential of combining FMs with other prediction tasks in wireless networks.

\subsection{Foundation Models for Other Prediction Tasks}

\begin{figure*}[!t]
\centering
\includegraphics[width=6in]{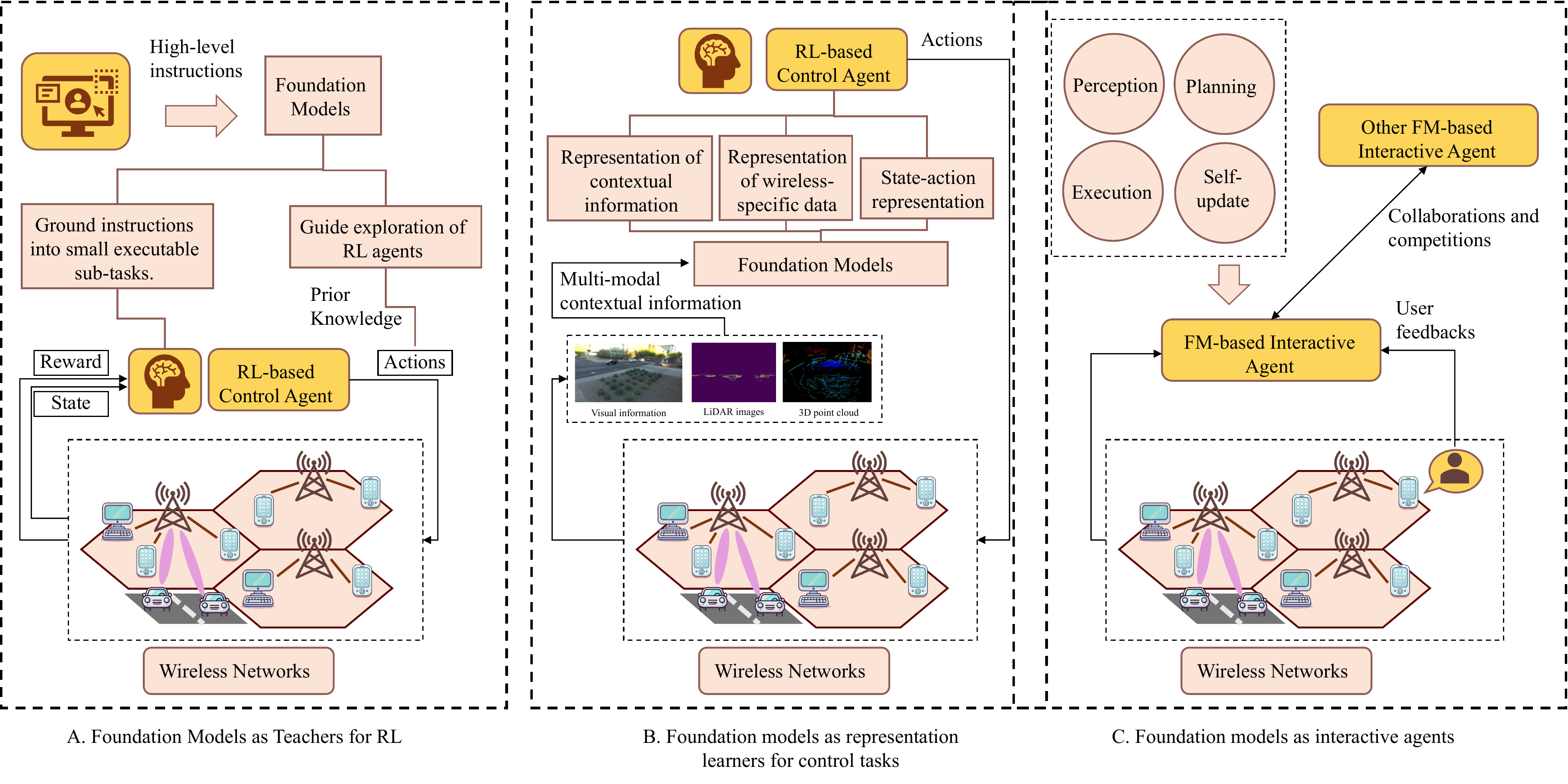}
\caption{Three ways to apply FMs to control tasks in wireless networks.}
\label{Control}
\vspace{-10pt}
\end{figure*}

In this subsection, we discuss the possibility of developing FMs or leveraging existing pre-trained FMs for four other fundamental prediction tasks in wireless networks: CSI prediction, radio link failure prediction, blockage prediction, and resource usage prediction.

\begin{itemize}
    \item \textbf{CSI prediction} CSI prediction refers to the process of predicting the future state of a wireless communication channel. It plays a fundamental role in facilitating massive multiple-input and multiple-output (m-MIMO) related design in wireless networks. CSI is usually high-dimensional structural data derived from the multi-path channel model, rather than simple single-dimensional data, which significantly increases processing complexity. This complexity has motivated research into leveraging FMs to handle the complex data patterns of CSI. \cite{ChannelStateInformationPredictionFor5G} proposes a method to convert CSI information into images, allowing it to be processed by convolutional neural networks (CNNs). In this method, raw channel data is segmented into small cells, each represented as a pixel in an image. Following this method, there is potential to process CSI information, represented as images, with vision FMs for prediction. In other words, pre-trained LLMs are fine-tuned for channel prediction tasks. For instance, \cite{LLM4CP} has built a channel prediction neural network based on a pre-trained GPT-2 model and fine-tuned it to predict the future downlink CSI sequence based on the historical uplink CSI sequence.
    Similarly, \cite{WiFo} proposed WiFo, a masked autoencoder-based FM for wireless channel prediction. By treating time- and frequency-domain prediction as a unified reconstruction task, it leverages self-supervised pre-training on diverse CSI data to generalize to new scenarios without fine-tuning. Experiments show consistent performance across multiple configurations and datasets for CSI prediction.
    Another instance is ChannelGPT \cite{ChannelGPTPaper}, a large pre-trained model designed to generate and predict wireless channel characteristics using multi-modal environment data. It supports fast adaptation through fine-tuning and enables tasks like CSI prediction and path loss estimation across diverse real-world scenarios.

    To go beyond task-specific models, FMs offer a scalable solution for CSI prediction by learning generalized spatio-temporal patterns from large and diverse datasets. Unlike conventional deep networks, FMs can capture long-range dependencies across antennas, subcarriers, and time steps, enabling robust prediction under varying channel conditions. When paired with auxiliary modalities (e.g., environment context), they also improve adaptability to real-world dynamics, making them well-suited for 6G scenarios where channels evolve rapidly and pilot overhead must be minimized.
    
    \item Radio link failure prediction: Radio link failure is a challenging problem in 5G networks given its negative impact on communication reliability and latency, especially for ultra-Reliable Low Latency Communications (uRLLC) traffic \cite{RadioLinkFailurePredictionIn5G}. By analyzing historical Key Performance Indicators (KPIs) and incorporating contextual data such as weather station observations, the radio link failure prediction process aims to proactively identify potential disruptions in radio link performance. Recent advancements in radio link failure prediction have leveraged transformer-based time series models for temporal feature extraction and have demonstrated better generalization capabilities compared to traditional long short-term memory (LSTM)-based methods \cite{AGeneralizedTransformerBasedRadioLink}. The success of such transformer-based methods provides a potential direction for developing generalized FMs specifically designed for radio link failure prediction.
    \item Blockage prediction: Millimeter-wave and terahertz networks rely on LOS links, and sudden link blockages pose significant threats to the reliability of the networks. Blockage prediction allows the network to anticipate future blockages and facilitates proactive handovers if a blockage is predicted. Existing literature has demonstrated the effectiveness of visual information in aiding blockage prediction, and FMs such as ViT have been employed in extracting essential information from multi-modal data \cite{VisionAided6GWireless, ghassemi2025generative}. Similarly, this approach can be extended to other modalities by utilizing multi-modal FMs to extract contextual information and enhance prediction accuracy.
    \item Resource usage prediction for network slices: Resource usage prediction for network slices involves determining the number of Physical Resource Blocks (PRBs) required by each slice. It can help with proactive slice provisioning in 5G networks. A case study in \cite{WirelessAgentLargeLanguageModel} shows the capabilities of LLMs to manage network slicing by analyzing information such as the total number of users in the cellular network and the status of network slices. This demonstrates the potential of FMs, especially LLMs, to interpret network status data and use these data to enhance resource usage prediction.
\end{itemize}

In summary, these examples illustrate two parallel and promising strategies for leveraging FMs in wireless prediction. The first is a domain-specific approach: building new FMs from scratch, such as LWM \cite{LargeWirelessModel} and WiFo \cite{WiFo}, pre-trained on massive wireless datasets (e.g., channel data) to understand the physical layer. The second is a general-purpose adaptation approach: leveraging existing, pre-trained FMs (like LLMs or ViT) as powerful reasoning engines or context encoders. These FMs analyze contextual information—from natural language reports and multi-modal sensor data—to provide high-level insights that improve prediction accuracy.

\section{Foundation Models for Control in Wireless Networks}
\label{s5}

This section explores the applications of FMs in control tasks within wireless networks. In traditional ML-based methods, intelligent agents learn strategies through reinforcement learning (RL), and the process includes four key components: observation, exploration, learning, and interaction with the environment. Building on this framework, FM applications for control tasks in wireless networks can be categorized into three main aspects, which are shown in \ref{Control}. The first aspect is to use FMs as teachers for unsupervised RL and provide guidance to RL agents during exploration and training. The second aspect is to use FMs as representation learners and leverage the capabilities of FMs to pre-process the observation information. The last aspect is to leverage agentic AI techniques and to transform FMs into interactive agents \cite{AgenticAIAutonomy}.

\subsection{Foundation Models as Teachers for RL}

In traditional RL frameworks, agents learn optimal strategies by exploring unknown environments through a trial-and-error process. While this approach is effective in many simulation scenarios, it poses significant challenges in complex environments or real-world applications requiring intricate decision-making. In such cases, agents usually require a long exploration time to find the relationship between actions and expected long-term rewards. This can lead to slow convergence, increased computational costs, and even suboptimal performance. Moreover, in wireless networks, certain actions taken during the exploration, if not handled carefully, may result in actions that degrade the performance of the network. 

Integrating FMs as teachers for RL can mitigate these concerns. Pre-trained on large-scale, diverse datasets, FMs possess common-sense knowledge and emergent capabilities. By leveraging the knowledge and capabilities, FMs can guide RL agents toward meaningful tasks and semantically relevant behaviors, effectively reducing the time required for exploration. In this context, FMs act as human-like teachers, providing direction and enhancing the exploration and training process of RL agents \cite{LiftUnsupervised}.

There are two primary approaches to leveraging FMs as teachers for RL-based control tasks in wireless networks, namely, \textbf{task planning and execution}, and \textbf{exploration guidance}. The first approach is to use FMs for task planning and execution. FMs—particularly LLMs—excel at high-level task planning. They can effectively ground instructions into well-defined tasks and decompose complex objectives into smaller, executable sub-tasks. Additionally, they can guide agents in prioritizing specific sub-tasks to achieve the overall objective \cite{FoundationModelsInRobotics}. 

For instance, in wireless networks, LLM can interpret a human operator’s instruction from e.g., “increase throughput” into three related sub-tasks: traffic steering, beamforming, and power allocation. These sub-tasks can then be completed respectively by RL-based models and be selectively activated by LLM according to the desired outcome \cite{LLMBasedIntentProcessing}.

Additionally, FMs specified in code-generation can generate executable code \cite{CODEXMODEL} that interacts with application programming interfaces (APIs) for task completion \cite{TaskmatrixAI}. By combining FMs with different capabilities, comprehensive guidance, and automation can be provided to enable RL-based agents to complete control tasks in a more feasible and automated way.

Recent research on FM-enhanced wireless networks has provided examples of using FMs for task planning. In \cite{LargeLanguageModelEnhancedMulti}, LLM-based planning agents are designed by employing the CoT  approach and plan-and-solve approach to decompose the original 6G communication task into a series of sub-tasks based on external retrieved communication knowledge. In \cite{LLMAgentsAs6GOrchestrator}, a tool library with APIs is constructed, and after fine-tuning, LLMs are able to decompose physical-layer tasks into action steps that correspond to APIs in the tool library.

The second approach is to use FMs to guide the exploration of agents and to avoid undesirable states. 
When humans learn strategies for control tasks, their exploration decisions are often based on their prior knowledge about the control task, allowing them to avoid expending effort on clearly irrational actions. In contrast, RL models are usually initialized randomly, which means they lack such prior knowledge. This absence of guidance can cause agents to take irrational actions that lead to undesirable states.

FMs, especially LLMs, can provide prior knowledge to RL agents for control tasks. The prior knowledge can be integrated either by pre-selecting optimal actions for RL models during exploration or by incorporating regularization-based constraints to the policy of RL models \cite{GenerativeAIInTheLoop}. In \cite{ReinforcementLearningWithFoundationPriors}, an RL from the foundation priors framework is proposed to improve the sampling efficiency of RL models by leveraging the policy, value, and success-reward priors from FMs. FMs have demonstrated their ability to explain the decision-making processes of deep RL models in 6G network slicing admission control tasks \cite{LeverageLLMsToExplain}. Moreover, through techniques such as fine-tuning and retrieving external domain-specific knowledge, FMs can effectively perform traffic analysis and network performance analysis \cite{MobileLLAMA}. 

These studies further validate the potential of using FMs as wireless network control-related prior knowledge sources. By leveraging this knowledge source, the RL agents can not only avoid undesirable states but also quickly understand new environments, achieving a jump-start with only a minimal number of iterations \cite{AnOverviewOfMachineLearning}. 

In addition, FMs usually contain competitive relational knowledge and have remarkable performance on open-domain question answering. So another possibility for FM-guided learning is to set their roles as knowledge bases for training RL models that do not have direct access to the task environment due to the cost concerns \cite{LanguageModelsAsKnowledge}.

\subsection{Foundation Models as Representation Learners for Control Tasks}

FMs can enhance the perception function of control agents and serve as powerful representation learners to extract universal representations for downstream control tasks.
To be more specific, they can be used as preprocessors or initializers for various perceptual components of control agents. FMs are capable of learning general representations from large, diverse datasets before fine-tuning a specific task. This is especially advantageous for control tasks in the wireless domain. In such control tasks, observations can include different data modalities and complex data patterns, but high-quality labeled datasets are often scarce. Leveraging FMs to initialize perception encoders enables an easier understanding of wireless communication environments with limited data. For example, in wireless networks, FMs can be used to initialize perception encoders for RL-based spectrum management. Traditional RL approaches require long-term exploration to learn interference patterns, leading to inefficient initial performance. By pre-training an FM on RF signal patterns, the RL agent can start with a prior understanding of the RF signal inputs, which can lead to faster convergence with limited interaction data.

As explained in Section \ref{s3}, pre-trained multi-modal FMs can extract latent states from various data types, including visual data, graph data, point cloud data, and radio data. For wireless control tasks that involve multi-modal inputs, these representations can provide key information that enhances control strategies and enables agents to perform more effectively. For instance, in \cite{MultiModalTransformerAndReinforcement}, a pre-trained multi-modal transformer is leveraged to process images, LiDAR data, and RADAR data to enhance the performance of a beam management task.

In addition to commonly used data modalities, FMs as wireless-specific learners can be designed to extract latent representations from wireless-specific data types. The benefit of pre-training such kinds of FMs is that they are not restricted to scenario-specific data and have fewer requirements for data quality. The primary objective of pre-training is to learn invariant representations from wide-scale data that can be effectively adapted to various downstream tasks. For example, FMs can be trained to adapt to CSI data through unsupervised learning. In \cite{CSIGPT}, a GPT model has been designed to capture high-dimensional features from received signals and extract CSI features. It was first pre-trained on synthetic CSI samples and then fine-tuned to adapt to specific application scenarios. This model can be combined with CSI-based control tasks, such as power control \cite{RobustTransmitPowerControl} and handover management \cite{ChannelChartingLocating}.

Another important representation that FMs can learn from radio signals and channel information in wireless networks is positional information \cite{BigAIModelsFor6G}. The positional information is widely used in various wireless control tasks, including beamforming in MIMO and mobility management. Traditional ML-based localization models may face challenges in learning robust representations due to imperfect channel estimates and temporal environmental variations. To fix this problem, FMs can be first pre-trained on various channel datasets to extract transferable representations, and the pre-trained FMs can serve as a base model for acquiring positional information in a robust way \cite{SelfSupervisedAndInvariant}.

An example is given in \cite{ott2024radio}, where a general transformer neural network is pre-trained on 5G channel measurements. A self-supervised learning method is adopted by randomly masking and dropping input information and making the model learn to reconstruct it. During the pre-training process, the model can implicitly learn the spatiotemporal patterns of the propagation environment and then be fine-tuned to extract positional information. 

Moreover, a more visionary and less applied idea is that FM techniques and architectures can be leveraged to enhance the optimization of sequential decision-making tasks by utilizing task-specific interactive data for representation learning. Unlike the previously discussed applications of FMs in wireless networks, this idea is specifically applicable to RL-based control tasks. FMs can learn the latent state or action space of an environment by grouping states and actions that yield similar transition dynamics to provide compact state action spaces for policy \cite{ContrastiveBehavioralSimilarity}. For example, in a power control task, different transmission power levels will lead to similar interference and throughput outcomes under certain channel conditions. Pre-trained with wireless network-related data, FMs can map these power levels to a more compact latent action space, allowing an RL agent to reduce the redundancy in state-action representations and make more efficient policy updates \cite{FoundationModelsForDecisionMaking}. Similarly, FMs can also learn sequence-level representations through temporal contrastive learning and such representations can help with the decision-making for MDP \cite{RepresentationLearningWithContrast}. For example, FM can be pre-trained to understand sequential CSI information and generate representations of adjacent CSI inputs for beam selection and resource allocation tasks \cite{ghassemi2025foundation}.

\subsection{Foundation Models as Interactive Agents}

The third method applying FMs for control in wireless networks is to leverage agentic AI techniques and use FMs as interactive agents \cite{AgentAI}. In control tasks, there are usually two types of interactions. The first type is the interaction between agents and the environment. The second type is the interaction between different agents, which is typically observed in multi-agent systems. Within the control system, FMs can serve two roles. They can either act as an interactive environment, providing feedback to the control agent, or directly function as an interactive control unit.

An intuitive way to model FMs as an interactive environment is through iterative prompting between FMs and the control system, and generating Markov decision process (MDP) based on the prompt history and FM outputs \cite{FoundationModelsForDecisionMaking}. This method can be used to initialize the control system or simulate potentially dangerous actions that cannot be tested in real-world environments. However, it requires high capabilities from FMs and is typically better suited to tasks requiring less domain-specific expertise, such as gaming and robotics.

Another way is to make FMs serve as an interactive environment by summarizing the changes and effects caused by actions in the real-world environment. They can change them into a form that can be easily utilized by the control system. This allows FMs to simulate complex environments to reduce real-world data dependence. Additionally, FMs can generate universal reward values for RL models, eliminating the need for manual reward design. For instance, TEXT2REWARD \cite{Text2Reward} introduced a framework to generate shaped, dense reward functions based on goals described in natural language. Similarly, \cite{Text2Reward} introduces an augmented reward acquisition method by learning a new reward function utilizing LLM outputs as auxiliary rewards. 

In current wireless network optimization, the RL agents typically require simulated or real-world environments to interact with, along with manually designed reward functions for complex, multi-objective tasks. Instead, FMs can be pre-trained on wireless network data, such as traffic patterns and communication protocols, to simulate network conditions and eliminate the need for task-specific manual reward engineering. For example, given an action such as allocating bandwidth to a specific user, the FM can interpret the decision in context and provide a reward score by reasoning over historical performance data, fairness policies, and energy efficiency constraints.

With the advent of user-centric wireless communication services in the 5G and 6G era, assessing network service satisfaction from the end-user's perspective is critical for evaluating performance \cite{UserCentricMethodologyFor}. However, user feedback in wireless networks is often non-numerical or implicit, making it difficult to directly use for optimizing network parameters in control tasks. Furthermore, user feedback, such as perception of network quality, can vary significantly between individuals. This makes it difficult to be uniformly represented by standard metrics. Translating user feedback into well-designed reward functions is a complex challenge that greatly impacts the performance of the control task. FMs, especially LLMs, have the ability to generalize information and can be fine-tuned with additional layers to translate task-specific reward functions effectively \cite{LLMBasedIntentProcessing}. For instance, the FM can translate natural language-based complaints such as "The call quality drops" or "The network is slow" into low reward scores. It can also be used to generate a reasonable reward score if different users have conflicting feedback, especially when there are multiple competing objectives. Instead of assigning equal weights to all objectives, FM can dynamically adjust reward values based on user priority levels and user profiles.

Other than using FMs as part of the interactive environment, a single FM or a combination of multiple FMs can also act as an interactive control unit and directly perform control tasks instead of traditional RL-based methods. Given the multi-modal perception ability, reasoning ability, and planning ability of some FMs, especially LLMs, they have demonstrated good performance in making open-ended decisions and performing control as independent agents \cite{PretrainedLanguageModelsForInteractive}. Using FMs as the core reasoning tool enables the agent to tackle a series of fundamental RL challenges, such as efficient exploration and limited pre-scheduled skills, which traditionally require separate, vertically designed algorithms \cite{TowardsAUnified}. For instance, in \cite{Voyager}, an LLM-powered agent is designed to autonomously explore the Minecraft world, acquire diverse skills, and make discoveries without human intervention. The agent includes three key components, an automatic curriculum for open-ended exploration, a skill library for storing behaviors, and an iterative prompting mechanism to generate code as actions. This design shares similarities with  RL in its focus on exploration and interaction. However, leveraging a fundamental knowledge base and emergent capabilities of LLMs, this approach shows a higher level of intelligence and self-evolving capability.
A recent example of this paradigm is KlonetAI~\cite{klonetai}, which demonstrates how general-purpose FMs, fine-tuned with ICL, can serve as AI agents for automating wireless network management. By interpreting human language intents, KlonetAI performs control and configuration of communication systems without needing extensive labeled training data or task-specific architectures. This showcases how prompting-based methods can be used to create adaptable and generalist network controllers from pretrained LLMs.

In recent wireless communication research, there have been some similar examples where FMs and LLMs are considered to act as control agents. In \cite{WhenLargeLanguageModel}, LLMs with distinct roles are distributed across mobile devices and edge servers to collaboratively perform decision-making tasks by interacting with the environment through text, API tools, and embodied actions. \cite{WirelessAgentLargeLanguageModel} designs an LLM-empowered agent that is capable of perception, memory storage, action-taking, and strategic planning. In network slicing management tasks, the agent achieves a lower resource occupancy rate while supporting a greater number of users compared to traditional methods. The FM-enhanced wireless network management agent can translate high-level operator intents into configurations and actions, enabling dynamic and responsive management without the need for manual intervention.

The application of FMs as interactive agents can also be extended to complex multi-agent systems, leveraging the intelligence of FMs to handle the collaboration and competition between different agents \cite{EnablingMobileAIAgent}. FM-enhanced agents have demonstrated the ability to simulate human-like strategic interactions and contribute to advancing game theory knowledge in gaming and social problems \cite{Alympics, GameTheoreticLLM}. FM-enabled game theory also has many potential applications in wireless networks, particularly in areas such as distributed spectrum and interference management and resource allocation. The limited availability of wireless communication resources and the presence of interference often result in competition among network users. FM-enabled game theory provides an analytical framework that enables users to  model and predict the decision-making processes and strategy selections of others, leading to more efficient and adaptive network management solutions \cite{GenerativeAIForGameTheory}.

Inspired by the idea of decision transformer and the FMs for prediction, there is also a possibility of developing FMs specific for decision-making and control tasks based on the transformer architecture. The decision transformer mitigates the issues of low sample efficiency and poor generalization ability of traditional RL models that usually require starting each new training process from the beginning. In \cite{DecisionTransformerForWirelessCommunication}, the decision transformer-based method has shown strong performance in wireless resource management tasks across different scenarios. As the model size increases, the transformer-based control model shows improved generalization over a variety of tasks and adapts fast to new tasks. These advancements indicate the potential for the agent to achieve AGI-level capabilities in control tasks in the near future \cite{LargeDecisionModels}.


\section{Developing Wireless-specific Foundation Models: Datasets}
\label{s6}

In this section, we provide an overview of wireless communication-related datasets suitable for developing wireless-specific FMs. Specifically, we categorize these datasets into three types: wireless traffic datasets, RF datasets, and datasets of various sensing modalities. Each category is introduced separately below.

\vspace{-5pt}
\subsection{Wireless Traffic Datasets}

\begin{table*}[h!]
\centering
\caption{Wireless Traffic Datasets}
\renewcommand\arraystretch{1.6}
\resizebox{2\columnwidth}{!}{
\begin{tabular}{|c|c|c|c|c|c|c|}
\hline
\textbf{Name} & \textbf{Date} & \textbf{Location} & \textbf{Data collected at} &  \textbf{Technology}  & \textbf{Granularity} & \textbf{Applications} \\ \hline

Telecom Italia \cite{traffic_milan} & 2014 & Milan, Italy & Base Station & 4G & Ten minutes  &  \makecell{Traffic flow prediction \\ Urban mobility analysis \\ Smart city planning \\ Social network analysis} \\ \hline

City Cellular Traffic Map \cite{c2tm} & 2012 & China & Base Station & 3G & One hour & \makecell{Spatiotemporal dependence analysis \\ Traffic flow prediction} \\ \hline

Cellular Traffic Analysis \cite{cell_traffic_analysis} & 2019 & \makecell{Simulated \\ (WireShark)} & UE (Android devices) & 4G & Ten seconds & \makecell{Traffic prediction \\ Energy Efficiency} \\ \hline

Beyond Throughput 4G \cite{beyond4g} &  2018 & Ireland & UE & 4G & One second & \makecell{Throughput prediction \\ Mobility Analysis} \\ \hline

Beyond Throughput 5G \cite{beyond5g} &  2020 & Ireland & UE & 5G & One second & \makecell{Throughput prediction \\ Mobility Analysis} \\ \hline

LTE dataset in NYC \cite{LTE_NYC} & 2019-2020 & New York, USA & UE & 4G & One second & \makecell{Bandwidth prediction \\ Handoff management} \\ \hline

UT Mobile Net Traffic \cite{utmobilenettraffic2021} & 2019 & Austin, USA & UE & WiFi & NA & \makecell{Application classification \\ Activity classification} \\ \hline

NanJing Dataset \cite{nanjing} & 2020 & Nanjing, China & Roadside units & IoV & NA & \makecell{Task offloading \\ Edge computing} \\ \hline

5G dataset \cite{real5g} & 2022 & South Korea & UE & 5G & NA & \makecell{Traffic analysis \\ Traffic prediction} \\ \hline

China Unicom One Cell \cite{unicom} & 2016-2017 & China & Base Station & 4G & Five minutes & Traffic prediction \\ \hline

Shanghai Telecom \cite{shanghai} & 2014 & Shanghai, China & BS/UE & 4G & NA & \makecell{Mobile edge computing \\ Traffic prediction} \\ \hline

CIKM 21 dataset \cite{cikm} & 2020 & Taiwan & UE & 4G & Five minutes & Traffic prediction \\ \hline

6G-PATH RAN Traffic \cite{6gdataset} & 2025 & Europe & RAN/BS & 6G & NA & \makecell{AI-RAN \\ Traffic prediction} \\ \hline

\end{tabular}}
\label{tab:datasets_traffic}
\vspace{-10pt}
\end{table*}

Wireless traffic datasets are essential for understanding and optimizing the performance of modern communication networks. These datasets capture diverse traffic patterns, user behaviors, and network conditions, providing valuable insights for tasks such as traffic prediction, resource allocation, and network optimization.
Table \ref{tab:datasets_traffic} provides a summary of the available wireless traffic datasets suitable for training FMs that utilize wireless traffic as input data. The details of these datasets are provided below:

\begin{itemize}
    \item Telecom Italia \cite{traffic_milan}: The Telecom Italia dataset includes telecommunication activity data collected from Milan and the Province of Trentino, covering the period from November 2013 to January 2014. During data collection, the site is divided into a 100x100 grid, with each cell representing a 235x235 meter area. The dataset captures Short Message Service (SMS), calls, and internet activities in each grid every ten minutes. 
    \item City Cellular Traffic map \cite{c2tm}: The City Cellular Traffic map dataset was collected in 2012 in a medium-sized city in China, and it includes data from 13,269 BSs. It records hourly information such as the number of mobile users, transferred packets, and bytes, along with BS locations. 
    \item Cellular traffic analysis \cite{cell_traffic_analysis}: The Cellular traffic analysis dataset consists of 48 days of real cellular traffic data collected on the user side using tools like Wireshark. The collected features include uplink/downlink packet counts, packet sizes, and communication protocols. 
    The dataset has a default time granularity of ten seconds.
    \item Beyond Throughput 4G \cite{beyond4g}: The Beyond Throughput 4G dataset includes 135 real-world 4G LTE traces collected from two major Irish mobile operators across various mobility patterns (static, pedestrian, car, bus, and train). Each trace averages 15 minutes, with throughput values ranging from 0 to 173 Mbps sampled at one-second intervals. The data is collected using the G-NetTrack Pro app and provides KPIs such as throughput, signal strength, channel quality indication (CQI), GPS coordinates, velocity, and handover events. 
    \item Beyond Throughput 5G \cite{beyond5g}: Similar to the Beyond Throughput 4G dataset \cite{beyond4g}, the Beyond Throughput 5G dataset includes real-world 5G traces collected from a major Irish mobile operator under two mobility patterns (static and driving) and two application scenarios (video streaming and file download). 
    \item LTE dataset in NYC \cite{LTE_NYC}: The NYC LTE dataset consists of approximately 30 hours of uninterrupted 4G traces collected across five fixed public transportation routes (subway and bus) using NetMonitorPro and iPerf tools. It includes multivariate features such as bandwidth, signal strength, frequency band, handoff events, and device speed.
    \item UT Mobile Net Traffic \cite{utmobilenettraffic2021}: The UT Mobile Net Traffic dataset contains over 21 million packets from 29 hours of mobile traffic, collected in a controlled environment using an automated platform. It includes traffic from 16 popular mobile applications with both application and activity-level labels. The dataset captures flow-based features from packet headers, such as packet size, direction, and timestamps, while ensuring privacy by excluding payload data. 
    \item NanJing dataset \cite{nanjing}: The NanJing dataset consists of real-world ITS social media data collected from 436 Roadside Units (RSUs) in Nanjing, China. It includes the geographic locations of RSUs and over 160 million vehicular social media service requests recorded over 30 consecutive days. 
    \item 5G dataset \cite{real5g}: The 5G dataset contains 328 hours of real 5G traffic traces collected from a South Korean mobile operator, covering diverse applications like video streaming, gaming, and video conferencing. It includes packet header information and timestamps.
    \item China Unicom One Cell \cite{unicom}: The China Unicom One Cell dataset includes 17 months of 4G mobile network traffic data from a single cell of China Unicom. The data is aggregated from Call Detail Records (CDR) into five-minute intervals, resulting in 140,256 data points. Each data point includes a timestamp and the total traffic volume (in bytes) for the five-minute period. 
    \item Shanghai Telecom \cite{shanghai}: The dataset provided by Shanghai Telecom, contains over 7.2 million records of mobile internet access through 3,233 BSs by 9,481 mobile phones over a six-month period. The dataset includes six key parameters: Month, Date, Start Time, End Time, BS Location (longitude and latitude), and User ID. 
    \item CIKM 21 dataset \cite{cikm}: The dataset contains six months of cellular traffic data from six road intersections, aggregated in five-minute intervals. It includes International Mobile Equipment Identity (IMEI) quantities to represent user mobility and captures spatial-temporal correlations.
    \item The 6G-PATH RAN Traffic Dataset \cite{6gdataset} provides real-world RAN traffic measurements collected from a 6G experimental testbed developed within the 6G-PATH project. The dataset includes time-series traffic and resource utilization statistics observed in a smart city connected and sensing use case. By offering real RAN-level traces rather than synthetic data, the dataset enables data-driven evaluation of AI-native network management, traffic prediction, and intelligent resource allocation techniques envisioned for future 6G systems.
\end{itemize}

\begin{table*}[h!]
\centering
\caption{Radio Frequency Datasets}
\renewcommand\arraystretch{1.2}
\resizebox{1.7\columnwidth}{!}{
\begin{tabular}{|c|c|c|c|c|}
\hline
\textbf{Dataset Name} &
  \textbf{Date} &
  \textbf{Data Collection} &
  \textbf{Signal} &
  \textbf{Application scenario} \\ \hline
RFF Dataset \cite{zhang2023radio} &
  2020 &
  \begin{tabular}[c]{@{}c@{}}Indoor environment\\ with a transmitter and\\ a receiver\end{tabular} &
  \begin{tabular}[c]{@{}c@{}}Real-world RF signals,\\  IQ samples\end{tabular} &
  RF fingerprint identification \\ \hline
\begin{tabular}[c]{@{}c@{}}ADS-B \\ Dataset \cite{adsb}\end{tabular} &
  2019 &
  \begin{tabular}[c]{@{}c@{}}Based on a special\\  aeronautical monitoring \\ system called ADS-B\\ in outdoor environment\end{tabular} &
  \begin{tabular}[c]{@{}c@{}}Over-the-air\\ real-world\\  ADS-B signals,\\ IQ samples\end{tabular} &
  Radio signal recognition \\ \hline
\begin{tabular}[c]{@{}c@{}}RFMLS \\ Dataset \cite{darpa_rfmls_2017}\end{tabular} &
  2018 &
  \begin{tabular}[c]{@{}c@{}}Real wireless devices\\  in both indoor and\\  outdoor environment\end{tabular} &
  \begin{tabular}[c]{@{}c@{}}Signals from different\\ communication\\ technologies, IQ samples\end{tabular} &
  Modulation recognition \\ \hline
\begin{tabular}[c]{@{}c@{}}RFMLS-NEU \\ Dataset \cite{rfmls_neu}\end{tabular} &
  2018 &
  \begin{tabular}[c]{@{}c@{}}Wireless transmissions\\  from over multiple devices\\ captured in the wild\end{tabular} &
  \begin{tabular}[c]{@{}c@{}}ADS-B signals and\\ WiFi signals, IQ samples\end{tabular} &
  RF fingerprint identification \\ \hline
\begin{tabular}[c]{@{}c@{}}RadioML \\ Dataset \cite{radioML}\end{tabular} &
  2016 &
  \begin{tabular}[c]{@{}c@{}}Simulated RF signals\\  with varying SNRs\end{tabular} &
  Synthetic I/Q samples &
  Modulation classification \\ \hline
\end{tabular}%
}
\label{tab:datasets_rf}
\vspace{-10pt}
\end{table*}

\vspace{-10pt}
\subsection{Radio Frequency Datasets}

In this subsection, we introduce the RF datasets. These datasets capture diverse characteristics of the RF signal, enabling the study of device identification, signal recognition, and communication safety. The details of these datasets are outlined below:

\begin{itemize}
    \item RFF dataset \cite{zhang2023radio}: The RFF dataset is designed to advance the radio-frequency fingerprinting (RFF) research targeting device identification and safety in wireless communication. It is acquired using an LTE-based communication system and employs ten software radio peripherals as transmitters and receivers. The dataset has signals captured in both LOS and non-LOS channel environments for diversity. Each sample comprises 30,720 data points, with 400 samples for each transmission. 
    \item Automatic Dependent Surveillance-Broadcast (ADS-B) dataset \cite{adsb}: The ADS-B dataset is a collection of ADS-B signals designed to support DL research in radio signal recognition. It includes 426,613 long signals and 167,234 short signals from thousands of airplane categories, captured using an automatic collection and labeling system in real-world environments.
    \item Radio Frequency Machine Learning Systems (RFMLS) dataset \cite{darpa_rfmls_2017}: The RFMLS dataset is designed to advance ML applications in the RF domain. It provides a rich training set of RF signals to enable systems to identify and classify known and unknown waveforms. The dataset supports key tasks like feature learning, saliency detection, and waveform synthesis, addressing challenges in crowded RF environments. 
    \item RFMLS-NEU dataset \cite{rfmls_neu}: The RFMLS-NEU dataset is based on the DARPA's RFMLS program and contains more than 11 terabytes of live RF signal sample data such as Wi-Fi and ADS-B transmission from over 10,000 devices. It provides complex-valued I/Q samples and metadata.
    \item RadioML \cite{radioML}: RadioML dataset includes two key Versions, RadioML 2016.10A and RadioML 2018.01A. It is a collection of synthetic RF signals in the I/Q signal form with different modulation types and varying SNR conditions. 
\end{itemize}

\vspace{-10pt}
\subsection{Sensing Modality Datasets}

\begin{table*}[h!]
\centering
\caption{Sensing Modality Datasets}
\renewcommand\arraystretch{1.6}
\resizebox{2\columnwidth}{!}{
\begin{tabular}{|c|c|c|c|c|c|c|}
\hline
\textbf{Dataset Name} & \textbf{Date} & \textbf{Location} & \textbf{Environment} &  \textbf{Sensing modalities} & \textbf{Technology}  & \textbf{Applications} \\ \hline

WALDO \cite{nist_waldo} & 2021 & Simulated & Indoor & RADAR & IEEE 802.11ay & Localization \cite{blandino2022tools} \\ \hline

RADAR for comm \cite{RADARcom} & 2020 & Simulated & Outdoor & RADAR & mmWave & Sensing assisted beamforming \cite{RADARcom} \\ \hline

LiDAR for comm \cite{LiDARcom} & 2019 & Simulated & Outdoor & LiDAR & mmWave & Sensing assisted beamforming \cite{LiDARcom} \\ \hline

ViWi \cite{alrabeiah2020viwi} & 2019 & Simulated & Outdoor & Vision & mmWave & \makecell{Beam tracking \cite{alrabeiah2020viwi, tian2021vision} \\ Blockage prediction \cite{VisionAided6GWireless}} \\ \hline

Multi-camera \cite{lin2024multi} & 2024 & Simulated & \makecell{Outdoor \\ C-V2X} & Vision & mmWave & \makecell{RSU selection and \\ beam searching \cite{lin2024multi}} \\ \hline

NEU \cite{NEU} & 2022 & Boston, USA & \makecell{Outdoor \\ C-V2X} & \makecell{LiDAR \\ GPS} & mmWave & Beam Prediction \cite{NEU} \\ \hline

FLASH \cite{flash} & 2022 & Boston, USA & \makecell{Outdoor \\ C-V2X} & \makecell{Vision \\ LiDAR \\ GPS} & mmWave & \makecell{Beam Selection \cite{gu2023meta} \\ Sector prediction \cite{flash}} \\ \hline

e-FLASH \cite{eflash} & 2022 & Boston, USA & \makecell{Outdoor \\ C-V2X} & \makecell{Vision \\ LiDAR \\ GPS} & mmWave & \makecell{Beam Selection \cite{eflash}} \\ \hline

DeepSense 6G \cite{alkhateeb2023deepsense} & 2021-2024 & Arizona, USA & \makecell{Outdoor (S1-41) \\ Indoor (S42-44)} & \makecell{LiDAR \\ Vision \\ RADAR \\ GPS \\ ISAC} & mmWave & \makecell{Beam Prediction \cite{DeepSenseChallenge} \\ Blockage Prediction \cite{charan2022computer} \\ Modality generation \cite{farzanullah2024generative}} \\ \hline

M3SC dataset \cite{m3sc} & 2023 & Simulated & Outdoor & \makecell{LiDAR \\ Vision \\ RADAR \\ GPS} & mmWave & \makecell{Channel modelling \cite{peng2024multimodal, huang2024scatterer} \\ CSI Learning \cite{zhang2024synesthesia}} \\ \hline

DISC \cite{disc} & 2023 & Real & Indoor & ISAC & IEEE 802.11ay & Activity recognition \cite{mazzieri2024attention} \\ \hline

Sensiverse \cite{sensiverse} & 2023 & 25 cities & Outdoor & ISAC & mmWave & Environment reconstruction \cite{10695938} \\ \hline
 
\end{tabular}}
\label{tab:datasets_sensing}
\vspace{-10pt}
\end{table*}

Sensing modality datasets are critical for advancing research in joint sensing and communication systems, which combine environmental sensing with wireless communication. These datasets encompass a variety of sensing modalities, such as LiDAR, RADAR, vision, and multi-modal data, enabling the development of intelligent wireless networks.
Table \ref{tab:datasets_sensing} summarizes the available sensing modality datasets that can be used for training multi-modal FM specific for wireless networks. The details of these datasets are provided below:

\begin{itemize}
    \item WALDO \cite{nist_waldo}: The dataset was designed to facilitate the development and evaluation of algorithms for communication and sensing within the 60 GHz frequency band. It comprises synthetically generated indoor mmWave channels between a MIMO transmitter and MIMO receivers, with multiple targets moving within the environment. The number of targets, their velocities, and trajectories are randomized throughout the dataset. Additionally, the dataset includes noisy received IEEE 802.11ay channel estimation fields. 
    \item RADAR for communications \cite{RADARcom}: This dataset is developed to validate the proposed RADAR-assisted mmWave beamforming scheme, in which the frequency-modulated continuous wave (FMCW) RADAR and communication systems operate at distinct frequencies. They configure the communication links at 73 GHz and RADAR at 76 GHz.
    \item LiDAR for communications \cite{LiDARcom}: In this dataset, BlenSor, SUMO, and Wireless InSite were used to generate realistic V2I mmWave communications. A Python orchestrator converts SUMO outputs (e.g., vehicle positions) and coordinates LiDAR and ray-tracing simulations for paired data generation.
    \item ViWi \cite{alrabeiah2020viwi}: The ViWi datasets offer co-existing visual data and parameterized wireless channel information, with sensors exclusively mounted on the RSU.
    \item FLASH \cite{flash}: Federated learning for automated selection of high-band mmwave sectors (FLASH) is a multi-modal dataset for cellular vehicle-to-everything (C-V2X), where the vehicles are equipped with camera, LiDAR, and GPS. 
    \item e-FLASH \cite{eflash}: The e-FLASH dataset is a comprehensive multi-modal dataset that includes camera images, GPS, and LiDAR data, accompanied by RF ground truth. 
    The dataset encompasses seven distinct real-world vehicular network scenarios, covering both LOS and non-LOS conditions.
    \item DeepSense 6G \cite{alkhateeb2023deepsense}: The Deepsense 6G dataset is a comprehensive multi-modal dataset designed to support research and development in next-generation wireless communication systems. It includes a variety of modalities such as RF signals, LiDAR, RGB video and GPS location data, enabling a holistic approach to sensing and communication tasks. The dataset captures diverse real-world scenarios, including indoor and outdoor environments, urban and rural settings, and dynamic conditions with varying levels of mobility and interference. 
    \item M3SC \cite{m3sc}: The M3SC dataset is a multi-modal dataset for 6G research, integrating RGB images, depth maps, LiDAR, RADAR, mmWave waveforms, and channel data. Collected using simulation tools, it covers diverse scenarios like urban, suburban, and rural settings under varying weather and traffic conditions. M3SC supports the development of ISAC systems for applications such as autonomous driving and smart cities.
    \item DISC \cite{disc}: This dataset is designed for research in ISAC using mmWave systems. It provides CIR measurements from IEEE 802.11ay packets, capturing reflections from human activities in indoor environments. The dataset includes uniformly sampled CIR sequences for seven subjects performing activities like walking, running, sitting, and waving, as well as sparse CIR sequences with realistic Wi-Fi traffic patterns. 
    \item Sensiverse \cite{sensiverse}: The Sensiverse dataset is a large-scale ISAC resource with 15TB of channel data across four frequency bands (3.5GHz to 100GHz) and over 25 cities and 100 scenes. 
    
\end{itemize}

Datasets will play a crucial role in the development of FMs for wireless communication domain. The availability of diverse, high-quality datasets allows FMs to learn from a wide range of scenarios, modalities, and environmental conditions, ensuring their adaptability and robustness across various tasks.
From RF datasets to multi-modal sensing datasets, these resources provide the foundational knowledge required for pre-training and fine-tuning FMs, enabling them to generalize 
effectively and perform complex prediction and control tasks.

\subsection{Data Suitability and Limitations}

While the datasets summarized in Tables~\ref{tab:datasets_traffic}--\ref{tab:datasets_sensing} provide diverse modalities for pre-training wireless-specific FMs, several factors should be considered when selecting datasets for FM training. 
First, dataset scale varies significantly across categories. Synthesized RF datasets such as RadioML offer millions of labeled samples but lack real-world channel impairments, whereas collections like DeepSense~\cite{alkhateeb2023deepsense} or FLASH~\cite{flash} provide realistic multi-modal data at the cost of smaller dataset sizes and limited measurement diversity. 
Second, noise characteristics also differ. Synthetic datasets allow controlled SNR variation, while real CSI measurements or RF captures include hardware distortions and environmental artifacts that may mitigate overfitting to overly clean data. 
Third, labeling availability is highly task-dependent. RF fingerprint datasets include device-level labels, whereas beam prediction or ISAC datasets often require custom labeling pipelines tied to ray-tracing or GPS ground truth. 
Finally, dataset applicability to downstream tasks depends on scene type (vehicular, indoor, massive MIMO, ISAC) and frequency band, which must match the deployment scenarios envisioned for the FM.

\subsection{Preprocessing Considerations for Wireless-Specific Modalities}

Unlike traditional datasets used in vision or language FMs, wireless datasets contain physical-layer modalities such as complex-valued CSI, raw baseband in-phase and quadrature (I/Q) samples, radar range–Doppler signatures, and scatterer-level geometric channel descriptions. These modalities reflect the physics of electromagnetic wave propagation and sensing, and therefore require modality-specific preprocessing pipelines that differ significantly from tokenization in text or pixel normalization in image data.

For example, CSI tensors are typically indexed by antenna, subcarrier, and time. They are first converted from complex-valued matrices into magnitude–phase or real–imaginary representations to support real-valued neural network architectures~\cite{LargeWirelessModel}. Depending on the FM architecture, CSI can then be reshaped into 2D grids (for CNN-based encoders~\cite{Farz2508:Conditional}), projected into patch sequences (for transformer-style models), or aligned with other features (in cross-modal learning settings~\cite{Farzan_beam}). 
Similarly, raw I/Q samples must be normalized in amplitude, optionally clipped to manage dynamic range, and either passed as real-valued sequences or transformed into time–frequency spectrograms or constellation diagrams for RF learning tasks. Recently, the IQFM model~\cite{iqfm} demonstrated that foundational models can be trained directly on raw multi-channel I/Q tensors using contrastive self-supervised learning, bypassing the need for handcrafted feature extraction while enabling multi-task learning across modulation classification, RF fingerprinting, angle-of-arrival estimation, and beam prediction.

Radar sensing data, such as those in the M3SC~\cite{m3sc} or DISC~\cite{disc} datasets, undergo additional signal processing such as windowing, Fourier transforms, and range–Doppler map extraction to form 2D or 3D representations. These structured radar outputs can then be directly fused with LiDAR or camera frames in multi-modal FMs for localization, sensing-assisted communications, or blockage prediction.

Overall, these preprocessing steps not only preserve the inherent spatio-temporal and complex-valued structure of wireless data, but also determine the choice of encoder architecture, positional encoding scheme, and modality fusion strategy in wireless-specific FMs. As such, careful preprocessing is a prerequisite for unlocking the full potential of raw RF data within FM pipelines.

\begin{figure*}[!t]
\centering
\includegraphics[width=5.3in]{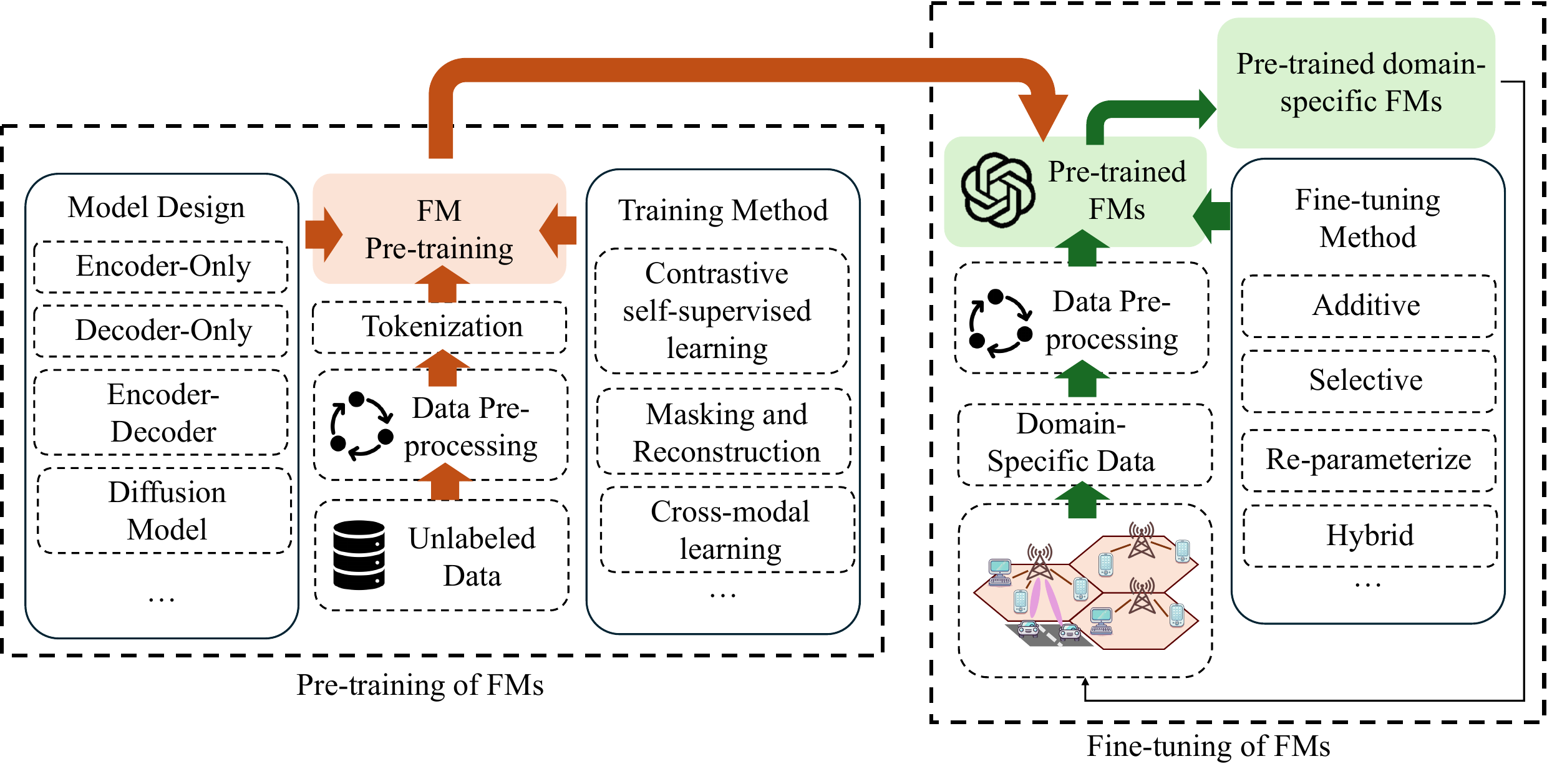}
\caption{The workflow of developing a domain-specific FM.}
\label{FMtraining}
\vspace{-10pt}
\end{figure*}

\section{Developing Wireless-specific Foundation Models: Methods}
\label{s7}

In this section, we discuss the methodologies for developing wireless-specific FMs. The workflow of developing a domain-specific FM is shown in Fig. \ref{FMtraining}. We first discuss two primary approaches for FM development: pre-training a model from scratch or fine-tuning an existing FM to incorporate wireless-specific knowledge. Additionally, federated learning is a special technique that can be employed to enhance the efficient development of wireless-specific FMs, and it is also discussed in this section.

\vspace{-5pt}
\subsection{Pre-training Foundation Models for Wireless Networks}

The goal of FM pre-training is to transform a predefined model architecture and a large, diverse wireless-specific dataset into a highly capable, generalized FM. This subsection explores key aspects of the pre-training process, including model design, tokenization strategies, and training strategies.

\subsubsection{Model design}

Transformer-based architecture is widely used in the design of FMs and is considered as one of the leading choices for model structure selection \cite{AttentionIsAllYouNeed}. The architecture is built based on an encoder-decoder framework and incorporates the attention mechanism as a core component of its design. The role of the encoder is to transform a sequence of input tokens into a fixed-length vector representation, while the decoder uses the latent representation to generate the corresponding output sequence. The encoder and decoder are jointly trained to maximize the conditional log-likelihood of the output given the input. 

Advancements have been made to the original transformer architecture over time, which stem from the observation that different structures can be beneficial for different tasks and applications.
As a result, transformer models are now commonly categorized into three classes: encoder-only transformers, decoder-only transformers, and encoder-decoder transformers. 

The pre-training of encoder-only transformers is typically performed by masking tokens in the input sentence and training the model to reconstruct those tokens. As a result, such models are good for tasks that require a deep understanding of input information. In contrast, decoder-only transformers are pre-trained to predict the next token in a sequence based on previous tokens. They are particularly well-suited for content-generation tasks. Encoder-decoder transformers combine the properties of both encoder-only transformers and decoder-only transformers and are more suited for tasks that involve generating new contents depending on a given complex input \cite{TransformerModelsAnIntroduction}. 

The selection of specific model structures for wireless-specific FMs depends on the nature of the task. For instance, for numerical time-series prediction tasks, decoder-only transformers are more suitable since the input data is relatively straightforward, and the focus is on generating predictions \cite{ADecoderOnlyFoundationModel}. However, when dealing with RF signals, which require a higher degree of information understanding, encoder-only or encoder-decoder transformers are more appropriate \cite{LargeWirelessModel}.

Diffusion models have also gained popularity in the design of FMs and can be another choice for the development for developing wireless-specific FMs for generative tasks \cite{CDDMChannelDenoising}. Diffusion models are trained by gradually adding noise to data, such as images until they become completely distorted. Then, the model learns to reverse this process by removing the noise to reconstruct the original data. Through this approach, diffusion models are trained to capture and understand the latent space that the data represent and to gain the ability to generate the data accurately \cite{DiffusionModels}. The diffusion mechanism itself is not dependent on the transformer architecture. However, many modern diffusion-based approaches have incorporated the transformer architecture as a backbone to achieve high-quality outputs in generation tasks \cite{ScalableDiffusionModels}. 

Recent research has also explored the use of non-transformer architectures to improve inference efficiency and reduce memory and compute requirements—particularly important in wireless applications where FMs may need to operate on edge devices or embedded systems. Among these, state space models (SSMs) have gained renewed attention as competitive alternatives to transformers for sequence modeling tasks.

For example, Receptance Weighted Key Value (RKMV) \cite{rkmv} is a hybrid RNN–Transformer architecture that retains the autoregressive capabilities of transformers while introducing recurrent mechanisms for token mixing. Unlike transformers that compute full-sequence attention, RWKV employs time-mixing and channel-mixing operations that scale linearly with sequence length, making it well-suited for low-latency, long-context applications on constrained hardware. Its minimal memory footprint and strong performance on long-sequence tasks make it a promising candidate for on-device FM deployment in wireless systems.

Another promising architecture is Mamba \cite{mamba}, a selective state space model designed for hardware-efficient sequence modeling. Mamba combines learned convolutional filters with gated state updates to capture long-range dependencies while avoiding the quadratic complexity of attention mechanisms. Its continuous-time formulation enables stateful, streaming inference, allowing Mamba to process incoming signal streams without reprocessing past data. These features are particularly advantageous in wireless signal processing tasks such as adaptive beamforming, mobility prediction, or real-time scheduling, where fast inference and low latency are essential.

Several other efficient FM architectures also target reduced computation and memory use during inference. For example, Linear Transformers~\cite{linear_transformer} and their variants (e.g., Performer~\cite{Performer} and FlashAttention~\cite{flashattention}) approximate attention with kernel-based or low-rank methods to scale to longer sequences with reduced compute. These models can be combined with modality-specific encoders (e.g., CNNs or GNNs) to enable efficient FM designs for RF sensing, beam prediction, and resource allocation.

Moreover, if FMs are designed for tasks that need to process specific types of input data, some additional model structures can be added as encoders to the model design. For example, when handling visual contextual information in wireless networks, adding convolutional layers enables FMs to understand and capture the correlations between neighboring pixels \cite{CvtIntroducingConvolutions}. Similarly, for tasks involving wireless network topologies represented as graph data, using GNNs allows the model to effectively capture intricate spatial relationships and dependencies between different network nodes \cite{RootCauseAnalysis}.

\subsubsection{Tokenization Strategies}

Implementing effective tokenization strategies allows FMs to handle diverse types of input data across various use cases. Tokenization is the process of dividing the input data into discrete units, called tokens and developing a digital, unique, and anonymous representation for each token. The primary goal of tokenization is to represent the input data in a manner that is meaningful for machines without losing its context. 

There are many pre-defined tokenization strategies. For instance, the natural language input can either be tokenized by characters, words or subwords \cite{BetweenWordsAndCharacters}. Similarly, the numerical input can be tokenized either at the single-digit level or as entire numbers, and for images, they are divided into overlapping or non-overlapping patches. The key point of selecting a tokenization strategy is to maintain a trade-off between feature expressiveness and information accessibility. Using smaller divisions can usually minimize the information loss, while larger divisions are better suited for extracting high-level features \cite{WhatMakesForGoodTokenizers}. 

Beyond the regular tokenization strategy selection rules, adaptive tokenization strategies can also be applied to include wireless-specific corpora into the tokenization design and augment the vocabulary of FMs with wireless-specific token sequences. These strategies can enhance the performance of FMs in domain-specific tasks \cite{EfficientDomainAdaptionOfLanguage}.

\subsubsection{Pre-training strategies}

Self-supervised learning is widely adopted for the pre-training of FMs. The benefits of self-supervised learning are that it enables FMs to learn from the unlabeled, raw data and thus avoids the cost of annotating large-scale datasets. Current self-supervised learning methods for training FMs are diverse. Some representative self-supervised learning methods that can be applied to wireless-specific FMs are introduced below:

\begin{itemize}
    \item Contrastive self-supervised learning: 
    Contrastive self-supervised learning trains FMs to make them bring the representations of augmented versions of the same sample closer together while pushing apart the representations of different samples. Similarity metrics are defined to measure how close two embeddings are \cite{AsurveyOnContrastive}. Some contrastive self-supervised learning methods have been shown to produce results comparable to the state-of-the-art supervised method, such as SwAV \cite{SwAV} and SimCLR \cite{SimCLR}.
    \item Masking and reconstruction: The masking and reconstruction-based self-supervised learning method is usually performed by masking certain parts of the input data and training the model to predict the masked parts. The idea is if the model is able to predict it successfully, it can learn the latent representation of the data. For instance, in \cite{RingMo}, an FM for remote sensing was trained by reconstructing the masked pixels in remote sensing images. This process enables the FM to learn general feature representations within the data distribution.
    \item Cross-modal learning: For tasks involving multi-modal data, cross-modal learning is an effective self-supervised learning method that leverages cross-modal information as an alternative source of supervision. This approach exploits mutual information between modalities to train robust and powerful feature representations for each modality \cite{CrossAndLearn}. For example, this method can be used to align historical traffic series data with network to build a multi-modal FM for wireless traffic prediction. The model can be trained by jointly embedding the multi-modal inputs into a shared latent space and minimizing a contrastive loss.
\end{itemize}

A key requirement of pre-training is to equip the model with the ability to generalize across a wide range of downstream tasks. While self-supervised learning on a large-scale, diverse dataset helps FMs avoid task-specific biases, applying an appropriate set of data augmentation strategies during the pre-training process can further enhance the generalization capabilities for downstream applications.

In the context of wireless-specific FMs, self-supervised learning must be adapted to the structural and physical properties of wireless data. Unlike natural images or text, wireless signals exhibit characteristics such as spatio-temporal correlation, channel reciprocity, frequency-selective fading, and sparsity in antenna–frequency grids. For example, contrastive SSL can exploit invariances in channel data by treating CSI samples collected from similar locations or propagation conditions as positive pairs, while treating those from dissimilar environments as negatives, thereby enforcing semantic alignment across channel states \cite{guler2025multi}. Masked reconstruction methods can be applied to CSI tensors or beamformed measurements by strategically masking parts of the time–frequency–space grid, enabling the model to learn structure-preserving interpolation and channel continuity, as demonstrated by recent channel FMs \cite{LargeWirelessModel}. Similarly, cross-modal SSL offers an opportunity to fuse wireless data with visual or environmental modalities (e.g., LiDAR, trajectory, or geographic context), enabling joint representation learning across sensing and communication domains \cite{ChannelGPTPaper}. By tailoring these SSL strategies to leverage the inherent domain properties of wireless signals, FMs can build more robust and transferable representations for downstream tasks such as traffic prediction, localization, and channel estimation.

A key motivation for adopting self-supervised learning in wireless FMs is its ability to reduce reliance on annotated datasets, which are especially scarce in multi-modal wireless settings. Since large quantities of raw CSI, mobility traces, and sensory data can be collected without labeling, SSL methods (e.g., masked reconstruction, contrastive learning, or cross-modal alignment) enable FMs to exploit structural properties of wireless signals without human annotation~\cite{LargeWirelessModel}. Combined with data-efficient techniques such as few-shot fine-tuning, synthetic-to-real domain adaptation, and parameter-efficient training, SSL provides a practical pathway for developing FMs even when labeled multi-modal data is limited.
\vspace{-5pt}

\subsection{Fine-tuning Foundation Models}

During the fine-tuning of FMs for wireless applications, several factors should be considered. First, the FM should incorporate wireless network-related knowledge while minimizing computational resource requirements, since the fine-tuning process is often performed under the severely constrained computational, memory, and power budgets of edge and end-devices. Second, the fine-tuning process should avoid the risk of overfitting domain-specific data and ensure the generalization ability of the FMs. Additionally, critical challenges such as catastrophic forgetting and model stability should be taken into consideration and carefully addressed during the fine-tuning \cite{LearnFromModelBeyond}.

There are several ways to perform fine-tuning to FMs. The most straightforward method is to directly fine-tune the parameters of FMs. Typically the FM parameters are fine-tuned in a supervised manner using labeled datasets. However, given the constraints mentioned above, full-model fine-tuning is often impractical. Consequently, parameter-efficient fine-tuning (PEFT) methods have been designed to mitigate these computational and storage overheads.

These methods can be divided into four categories: additive PEFT, selective PEFT, re-parameterized PEFT, and hybrid PEFT. 
Additive PEFT means that the FM can be fine-tuned by modifying the model architecture and injecting new trainable modules or parameters. For example, \cite{ImprovingLanguageUnderstanding} adds an additional task-specific linear layer to the original FM architecture to generate the desired outputs based on task labels. Only the parameters of the newly added layer are fine-tuned. 
The adapter-based fine-tuning method is a special type of additive PEFT method that inserts small adapter layers within transformer blocks. Early versions of adapters, often referred to as serial adapters, were designed to be inserted sequentially within the layers of transformers \cite{ParameterEfficientTransfer}. On this basis, parallel adapters are designed to reorganize the adapter layers into a parallel side network and to further reduce the computational cost associated with adapter layers \cite{TowardsAunifiedView}.

Selective PEFT methods freeze the majority of the original FM parameters and update only a small selected portion of parameters. For instance, \cite{LessIsMoreSelective} proposes a PEFT method by choosing only a small subset of layers for fine-tuning based on a greedy selection strategy.

Re-parameterized PEFT means that the FM architecture can be transformed to construct a low-rank parameterization and to achieve the goal of parameter efficiency during fine-tuning. The most widely recognized re-parameterized PEFT method is low-rank adaptation (LoRA). LoRA is developed based on the hypothesis that the change in weights during FM adaptation has a low intrinsic rank, and it is performed by freezing the pre-trained model weights and injecting trainable rank decomposition matrices into each layer of the transformer architecture \cite{LoRALowRank}. LoRA has proven to be a widely adopted and effective approach for domain adaptation tasks in FMs that can lead to strong performance across various applications \cite{LowRankAdaptationOfTimeSeries}.
Since performance varies by application, recent advancements have introduced hybrid PEFT approaches that combine diverse PEFT methods \cite{Unipelt, LLMadapters,ParameterEfficientFineTuning}.

Although PEFT reduces parameter counts, efficient deployment on the aforementioned heterogeneous edge hardware (ranging from CPUs to FPGAs) requires addressing further system-level intricacies beyond simple parameter reduction. This necessitates hardware-aware fine-tuning and advanced compression techniques \cite{marculescu2018hardware}.

Quantization, for example, is a critical technique in this context \cite{QuantizationSurvey}. It involves reducing the numerical precision of the model's weights and activations (e.g., from 32-bit floats (FP32) to 8-bit integers (INT8)). Post-Training Quantization (PTQ) is a simple method applied after fine-tuning, but it can lead to accuracy degradation. A more robust approach is Quantization-Aware Training (QAT), where the quantization process is simulated during the fine-tuning itself. This allows the model to adapt to the precision loss, preserving high accuracy \cite{QAT_Paper}. The true benefit is realized when the target hardware features specific tensor operators for low-precision arithmetic (e.g., on a mobile GPU or a custom ASIC), which drastically accelerates inference and reduces power consumption.

Furthermore, for tasks requiring more computation than a single edge device can handle, distributed techniques are necessary. This can involve distributed data parallelism across a cluster of edge servers (e.g., at gNodeB locations) or more complex model parallelism where parts of the model are split across devices. This becomes exceptionally challenging in wireless networks due to the heterogeneous nature of edge devices and unreliable network links, requiring adaptive scheduling and fault-tolerant algorithms \cite{Distributed_Edge_AI}. These techniques are distinct from, but complementary to, the Federated Learning discussed in Section VII-C, as they focus more on compute-resource sharing rather than data privacy.

In addition to fine-tuning the FM parameters, another efficient tuning method is to perform prompt-based fine-tuning and tune only the continuous prompts without altering the core architecture of FMs. For example, in \cite{ThePowerOfScale}, several additional trainable tokens, referred to as soft prompts, are added to the input text. Only the embeddings of these prompt tokens are optimized through end-to-end training while the pre-trained model remains fixed. Compared to parameter-based fine-tuning methods, prompt tuning is more storage-efficient. This is because it allows downstream tasks to use the original pre-trained model directly, without the need to create a task-specific copy of the entire model.

Other than the above-mentioned techniques, another technique that can be used for FM fine-tuning at a small cost is model editing. For instance, in \cite{FastModelEditing}, a collection of small auxiliary editing networks were proposed that can use a single input-output pair to make fast, local edits to a pre-trained model. Other than fine-tuning a pre-trained FM with a large domain-specific dataset, model editing is usually used to change some specific knowledge of an FM without adversely affecting the results of other inputs \cite{LearnFromModelBeyond}.

\subsection{On-Device and Distributed Inference}

Deploying large-scale FM in wireless networks is challenging due to the limited compute, memory, and energy resources of edge devices, along with the need for low-latency and privacy-preserving inference. Modern FM, such as LLMs, contain billions of parameters that typically require cloud-scale infrastructure. However, cloud-based inference increases bandwidth demand and raises latency and privacy concerns~\cite{edgeshard}. 
Running inference directly at the network edge—on BS, access points, or user devices—can mitigate these issues but remains constrained by the hardware capabilities of such devices.

On-device inference can be achieved through model compression and optimization. Llama.cpp~\cite{llama.cpp} demonstrated that 4-bit quantization drastically reduces memory usage, enabling 7B–13B parameter models to run on CPUs with modest performance loss. Techniques such as pruning and knowledge distillation further shrink model size~\cite{prima.cpp}, supporting private inference on commodity hardware. Optimized compilers like MLC-LLM~\cite{mlc-llm} generate hardware-specific executables for GPUs and NPUs, while AirLLM~\cite{airllm2023} streams model layers just-in-time, running 70B+ models with only 4~GB of GPU memory—making large-scale inference feasible on constrained devices.

Distributed inference frameworks enable multiple edge devices to cooperatively run large models by partitioning layers across nodes. Projects like Exo \cite{exo2024} and Distributed Llama \cite{dllama} connect heterogeneous devices in a peer-to-peer setup, sharing activations and balancing workloads to overcome single-device limits. Prima.cpp \cite{prima.cpp} further optimizes this via pipelined-ring parallelism and a heterogeneity-aware scheduler, achieving sub-second latency and up to 17× speedup on 70B-parameter models. Such collaboration demonstrates that powerful FMs can be efficiently deployed across distributed edge systems.

Recent work in edge AI explores distributing FM inference across wireless devices to improve efficiency and latency. ~\cite{zhang2025distributed} propose an over-the-air aggregation framework, exploiting the wireless channel’s superposition property to perform fast all-reduce operations and cut communication delay.
EdgeShard~\cite{edgeshard} optimizes model partitioning by jointly selecting devices and allocating model segments via dynamic programming, achieving up to 50\% lower latency and 2$\times$ higher throughput on Llama2 models. Galaxy~\cite{Galaxy} employs hybrid parallelism with tile-based overlap of communication and computation for heterogeneous edge clusters, reducing latency by 2.5$\times$.
LinguaLinked~\cite{lingualinked} links multiple smartphones into a cooperative cluster, using linear optimization and dynamic load balancing to enable fully on-device large-model inference without cloud support.

Another direction focuses on improving parallel execution for multi-device inference. TPI-LLM~\cite{tpi-llm} shows that tensor parallelism can outperform pipeline parallelism for single-user scenarios common in wireless applications, if communication is efficiently managed. It employs a sliding-window memory scheduler to stream model weights on demand and a star-topology all-reduce protocol to mitigate network latency. This design enables 70B-model inference across low-resource devices with only $\sim$3.1~GB memory per device, achieving over 80\% lower latency and 90\% lower peak memory use than single-device inference.
Automated scheduling has also been explored: \cite{automaticparallel} propose a self-optimizing scheduler that dynamically partitions Transformer models across heterogeneous edge devices based on resource availability and network conditions, ensuring balanced load and minimal response time in fluctuating wireless environments.

Overall, the integration of these strategies—from quantization and compiler optimizations for single-node efficiency to distributed partitioning and coordination for multi-node deployments—is enabling FM inference at the wireless network edge. Such advances allow powerful models to run locally for tasks like real-time signal analysis, network control, and user interaction without relying on remote cloud servers. This trend points toward future wireless networks composed of federated edge clusters collaboratively executing large models, combining high performance with strong privacy guarantees.

\vspace{-5pt}
\subsection{Privacy in Federated Learning-enabled Collaborative Foundation Model Development}

Federated learning enables the training, fine-tuning, and inference of FMs in a distributed manner, promoting the emergence of collaborative FM development. This collaboration benefits FMs in terms of both data availability and computational efficiency \cite{WhenFoundationModel}. On the one hand, training and fine-tuning FMs often face significant challenges in accessing sufficient high-quality public data. Federated learning can address this issue by allowing FMs to utilize abundant data from a wide range of sources without the need to collect user data directly. 

On the other hand, training large-scale FMs typically requires considerable computational resources. This creates a challenge for centralized training methods since a single server may lack the capacity to train the FM independently. Federated learning addresses this issue by enabling computation sharing and allowing participants to pool their computational power. In this way, the training process can be distributed and the burden on individual servers alleviated.

FL-enabled collaborative FM development is particularly suitable for wireless networks, where collecting and centralizing wireless-specific training data from diverse devices and vendors is challenging due to privacy regulations. The fundamental privacy benefit of FL stems from keeping raw data decentralized on user devices. However, FL is not a complete privacy solution; the transmission of model parameters or gradients to a central server can still be vulnerable to inference and inversion attacks, potentially leaking sensitive information from the local data. This privacy risk is coupled with high communication and computational costs. Transmitting large-scale FM parameters not only strains the wireless network (a high communication cost) but also significantly increases the privacy attack surface. Similarly, the high computational and storage demands for deploying FMs can make it infeasible to implement more complex, privacy-enhancing mechanisms (e.g., differential privacy or cryptographic aggregation) on resource-constrained devices \cite{wang2024data}.

Several existing studies propose concrete methods for developing federated learning-based frameworks for FMs. In \cite{Feddat}, a framework called Federated Dual-Adapter Teacher (FedDAT) was proposed to perform PEFT for FMs with federated learning. Specifically, it leveraged an adapter-based PEFT method and utilized a module comprising two parallel adapters. One adapter was kept frozen as a copy of the global adapter, while the other adapter was locally optimized for each client. \cite{PromptFL} proposed a federated prompt training framework for FMs and let distributed clients cooperatively train shared soft prompts based on very few local data.

Federated Learning-enabled collaborative FM development is particularly suitable for wireless networks, where collecting and centralizing wireless-specific training data from diverse devices and vendors is challenging due to privacy regulations. Despite the benefits, the combination of federated learning and FM in wireless networks also faces some challenges that need to be carefully addressed. A major challenge of combining federated learning with FMs is that model parameters or updates are usually required to be transmitted between clients and the central server, and this can lead to a high communication cost. Another challenge is that deploying FMs can bring high computational and storage demands, which typical network architectures may struggle to accommodate.

Recent studies explore how to apply federated FM in wireless networks and solve potential challenges.
In \cite{TheRoleOfFederated}, a hybrid training scheme for developing FMs over wireless networks was proposed to efficiently allocate the resources between central and distributed servers. With this scheme, the overall training process of FMs was split into two parts. In the first stage, the cloud or edge server would perform pre-training with the centralized public data. In the second stage, distributed clients perform fine-tuning based on local private data. \cite{FederatedFineTuning} proposed a split federated LoRA framework. It deployed the computationally intensive encoder of a pre-trained FM at the edge server while keeping the embedding and task modules at the edge devices. With this framework, the proposed framework only aggregates the gradient of the low-rank matrix, thereby reducing communication overhead.

\vspace{-5pt}
\section{Open Issues and Future Directions}
\label{s8}
While current research has shown that FMs offer numerous valuable applications in wireless networks, several critical challenges remain unresolved. Below, we highlight these key challenges and explore promising avenues for future research.

\vspace{-5pt}
\subsection{Development of Foundation Models for Wireless-Specific Modalities}

Wireless networks generate data of specialized modalities, such as network topologies, RF signals, CSI, ISAC data, and user traces. These data types often have high dimensionalities and are not directly compatible with pre-trained FMs designed for natural language inputs or vision inputs. Additionally, the information extracted from these modalities is often applicable across various wireless network-related tasks. As a result, there is a need to develop FMs to handle wireless-specific data modalities effectively. 

The datasets and methodologies for developing such FMs were discussed in Section \ref{s6} and Section \ref{s7}. Since wireless data are typically unlabeled, self-supervised learning is a promising approach for pre-training FMs on raw wireless data. \cite{LargeWirelessModel} can be seen as an initial attempt of developing such FMs and provides an approach that can be improved and extended to other wireless-specific modalities. These FMs are expected to extract latent representations from high-dimensional inputs and generate meaningful embeddings based on invariant characteristics of wireless data. By integrating additional layers and fine-tuning, these wireless-specific FMs can be adapted for diverse downstream prediction and control tasks. Furthermore, FMs trained on different data modalities can be combined and aligned using cross-modal learning techniques.

\vspace{-5pt}
\subsection{Empowering Pre-trained Foundation Models with Wireless-Specific Knowledge} 

Pre-trained FMs and LLMs have demonstrated advanced capabilities. However, current research on FMs for wireless networks primarily focuses on leveraging pre-trained models originally developed for NLP tasks. These FMs often lack wireless-specific knowledge and this limitation can result in suboptimal performance in wireless network-related tasks, such as task decomposition and action planning. Moreover, the applications of FMs in modeling the physical world of wireless communication is limited due to a lack of embodied experience and physical grounding. \cite{ArtificialGeneralIntelligenceNativeWireless,LargeMultiModalModels}. In the future, it is crucial to address this gap to enhance the applications of FMs in the wireless domain. 

One approach to overcome this challenge is prompt engineering, with techniques such as ICL and retrieval-augmented generation (RAG). These methods enable a cost-effective domain adaptation without modifying the FM parameters. However, prompt engineering may lead to inconsistent performance across different scenarios. In contrast, fine-tuning pre-trained FMs offers a more reliable and one-time solution with more stable performance \cite{PromptEngineeringOrFineTuning}. For example, \cite{MobileLLAMA} made an initial effort to incorporate wireless-specific knowledge into the LLAMA model through instruction fine-tuning.

A significant obstacle to fine-tuning FMs for wireless networks is the limited availability of labeled wireless data, and such data are always constrained by privacy concerns. To address this, future research should focus on designing privacy-preserving fine-tuning methods, using techniques such as federated learning and split learning, which can securely utilize decentralized data without compromising privacy.

Moreover, a promising research direction is the exploration of real-time model updating, where FMs adapt continuously to changing wireless environments such as user mobility or traffic variations. Recent studies, such as the device–edge cooperative fine-tuning framework proposed in~\cite{DeviceEdgeCooperativeFineTuning}, have shown the feasibility of adapting large models directly at the network edge without retraining from scratch.
Techniques such as online fine-tuning, streaming-based adaptation, or dynamic prompting could allow FMs to maintain performance under time-varying conditions, thereby enabling more responsive wireless systems.

\subsection{Exploring the Deployment of Foundation Models in Wireless Networks} 

Another practical challenge in applying FMs to wireless networks is their deployment. Deploying FMs typically requires substantial computational and storage resources, which vary significantly across different network nodes. While various deployment methods have been explored in previous studies, as discussed in Section \ref{s2}, many existing deployment frameworks lack the necessary details and thorough verification. This gap represents a promising direction for future research.

There are two specific problems that need to be discussed in future research on FM deployment. The first problem is how to achieve fast inference for large FMs. Potential solutions include employing parallelization techniques and batch processing of inputs or using parameter quantization to trade precision for speed and efficiency. The second problem is to address the high overhead costs associated with FMs under distributed deployment. In particular, it is crucial to determine how to allocate communication resources, such as bandwidth, effectively between signaling transmission and communication data transmission.

Additionally, integrating FMs across heterogeneous network layers, from terrestrial BSs to UAV relays, maritime communication nodes, and satellite systems, opens new opportunities for unified, multi-domain network intelligence. Such integration is central to the emerging vision of multi-layer 6G architectures, which combine Non-Terrestrial Networks (NTNs) with terrestrial Radio Access Networks (RANs) to provide seamless global coverage. The key challenges include designing FM architectures that can accommodate diverse link budgets, latency constraints, mobility profiles, and spectrum policies across layers.

Moreover, unlike traditional single-region deployments, real-world wireless environments demand models capable of transferring and adapting across differing propagation conditions and resource constraints. Recent research on hierarchical and federated FM design~\cite{Hierarchical} suggests promising directions. It leverages cooperative edge nodes, UAV-based relays, and low-earth-orbit satellites to collaboratively pre-train or fine-tune shared models across network tiers. This line of work may lead to FMs capable of learning cross-layer abstractions of channel dynamics, as well as enabling multi-RAT (Radio Access Technology) coordination, joint beamforming, or distributed cognition across the network stack.

\vspace{-5pt}
\subsection{Enhancing the Security of Foundation Models}

Another critical challenge associated with FMs is security. Security is a fundamental requirement in wireless communication. However, the integration of FMs into wireless networks may introduce new vulnerabilities that must be addressed.

On the one hand, FMs are typically trained on large-scale datasets, which increases the risk of including poisoned data. On the other hand, the transmission of model parameters or updates between network nodes introduces vulnerabilities to over-the-air attacks, further threatening system security. 
Unlike traditional ML models, FMs are usually over-parameterized with billions of parameters. This redundancy results in a sparse structure that attackers can exploit to embed hidden backdoors or triggers to attack FMs. These vulnerabilities pose significant risks to wireless networks, as they can lead to single points of failure in wireless communication systems. Therefore, future research is required to strengthen security measures and mitigate such attacks, thus building a trustworthy, FM-enhanced wireless communication environment.

\vspace{-5pt}
\subsection{Enhancing the Robustness of Foundation Models} 

Hallucination is another challenge when deploying FMs in wireless networks. Hallucination refers to the phenomenon where FMs may generate factually incorrect, nonsensical, or contextually inaccurate outputs. Such behavior can reduce the robustness of FMs and degrade the performance and reliability of wireless communication systems.

One approach to improving the robustness of FMs is to incorporate guardrails by introducing mechanisms and constraints during model development \cite{SafeguardingLargeLanguageModels}. For example, incorporating a diverse range of training data, applying data filtering techniques and applying robust training methods can improve stability and ensure more reliable performance. Another potential solution is to enhance FMs with wireless-specific knowledge through methods such as fine-tuning or RAG. By grounding FMs in domain-specific data, it is possible to reduce the overgeneralization of models and improve their task-specific performance. Moreover, integrating reasoning steps into the inference process can enhance the models' understanding of inputs and avoid the generation of conflicting outputs.

Exploring these strategies is important to ensuring that FMs can be deployed effectively in wireless networks, minimizing risks associated with hallucination and enhancing overall system robustness.

\vspace{-5pt}
\subsection{Developing On-device Small-Scale Foundation Models} 

Another promising direction for future research is to develop small-scale, on-device FMs that retain the essential capabilities required for wireless network-related tasks. On-device deployment allows FMs to adapt to tasks based on user interactions and to enhance personalization while preserving privacy. However,the current large size of FMs limits their practical use in wireless networks. According to existing research, FMs with fewer than ten billion parameters are typically ideal for operating effectively on various edge devices that are equipped with GPUs \cite{Qualcomm2023,TransformerLiteHighEfficiencyDeploymentLarge}. For real-time applications in wireless networks with stricter latency requirements, smaller FMs are required. FMs are typically over-parameterized and have large redundancy. This characteristic of FMs presents an opportunity to reduce their size and construct lightweight FMs for wireless network tasks.

Several techniques can be employed to reduce the size of FMs. For instance, knowledge distillation techniques can be employed to extract and transfer wireless network-related knowledge from large-scale FMs to small-scale FMs. Additionally, other model compression techniques, such as model pruning, quantization, and low-rank factorization, can further optimize the size and efficiency of FMs for on-device usage.

Going forward, hybrid on-device and cloud-based split inference architectures offer a promising way to balance energy, privacy, and performance in FM-enabled wireless systems. In this paradigm, lightweight model layers run locally on edge devices, while deeper layers are offloaded to nearby MEC or cloud servers. This reduces latency and energy consumption while preserving data privacy. Recent work, such as ~\cite{hybrid_add}, shows that adaptive split inference based on network conditions enables efficient real-time deployment of large models. Combining this approach with techniques like opportunistic offloading or knowledge distillation from larger edge-deployed models may further enhance scalability under resource constraints.

\subsection{Latency Considerations in FM-Enabled Wireless Systems}
Real-time operation is a critical requirement in many wireless networking tasks such as link adaptation, beamforming, and resource scheduling. While FMs offer strong generalization capabilities, their large computational footprint introduces significant inference latency, potentially making them unsuitable for latency-sensitive applications without appropriate system-level adaptations. As noted in~\cite{latency}, traditional cloud-based FM deployments often incur round-trip delays on the order of 50--200~ms due to network propagation and centralized inference bottlenecks. To mitigate this issue, recent approaches advocate moving inference closer to the network edge by leveraging Telco infrastructure such as multi-access edge computing (MEC) nodes or RAN-level processing, thereby reducing end-to-end latency to sub-10~ms scales. Such edge acceleration strategies, including model partitioning, caching, or split inference, are essential for enabling FM-driven intelligence in time-critical wireless applications while maintaining operational feasibility and responsiveness.

\subsection{Overcoming Data Scarcity for Multi-Modal FM Development}
A major practical challenge in developing wireless FMs is the scarcity of large-scale, high-quality, and well-labeled datasets, especially in the multi-modal setting where CSI, visual, LiDAR, mobility, and RF sensing data must be spatially and temporally aligned. Unlike NLP or CV domains, where web-scale datasets enable billion-parameter pre-training, wireless data collection requires specialized hardware, controlled environments, and often contains privacy-sensitive user information, making large annotated datasets difficult to obtain.

To address this limitation, several research directions are promising. First, self-supervised pre-training on raw, unlabeled wireless data can significantly reduce dependence on annotation, as demonstrated in emerging wireless-specific FMs~\cite{LargeWirelessModel,guler2025multi}. Second, synthetic or semi-synthetic dataset generation using ray-tracing engines and wireless digital twins can expand training diversity without requiring physical measurements. Wireless AI surveys note that data is often scarce, incomplete, and costly to acquire~\cite{datascarce}. Third, privacy-preserving data federation approaches such as federated or split learning across operators and testbeds can enable access to distributed datasets without sharing raw data~\cite{Hierarchical}. Finally, data-efficient training strategies, such as few-shot adaptation, parameter-efficient fine-tuning, or cross-modal knowledge transfer, represent a viable path for scaling FMs even when labeled multi-modal datasets remain limited.

\subsection{Towards Agentic AI for Autonomous Wireless Networks}

Agentic AI augments FMs with perception, memory, reasoning, and tool-use capabilities to create autonomous decision-making agents. Unlike static predictors, agentic systems operate in a closed loop: perceiving multi-modal states, reasoning over objectives, and executing actions~\cite{ReAct,Toolformer}. This paradigm is well-suited for wireless networks, where complex, coupled decisions regarding spectrum, power, and mobility require dynamic adaptation beyond static optimization.

Recent frameworks like ReAct~\cite{ReAct} and Toolformer~\cite{Toolformer} demonstrate how FMs can iteratively plan and invoke external tools. In wireless contexts, such agents can translate high-level operator intents into executable configurations, balancing conflicting objectives like throughput and energy efficiency~\cite{WirelessAgentLargeLanguageModel,WhenLargeLanguageModel}. Compared to traditional reinforcement learning, agentic FMs offer greater flexibility in handling open-ended goals and heterogeneous data.

However, deployment faces challenges, including real-time latency constraints, hallucination risks, and the difficulty of integrating with existing protocol stacks. Despite these hurdles, agentic AI represents a natural evolution from passive modeling to active decision-making, promising self-optimizing networks capable of operating with minimal manual intervention.


%
\vspace{-10pt}
\section{Conclusion}
\label{s9}
This work explores the potential of applying FMs to wireless networks. Specifically, it examines existing and prospective applications of FMs in processing multi-modal data within wireless networks and their role in enhancing the performance of both prediction and control tasks. Given that the feasibility of certain tasks relies on the assumption that pre-trained FMs contain wireless-specific knowledge, this work also discusses the datasets and methodologies required to develop wireless-specific FMs.

In addition to summarizing recent research advancements, this survey paper also provides a roadmap for the future integration of FMs into wireless networks. It highlights key challenges in this field and explores possible solutions and research directions. By presenting a comprehensive survey and outlining future opportunities, it provides a solid foundation for further research and encourages new explorations in this rapidly evolving field.

\section*{Acknowledgment}
This work has been supported by MITACS and Ericsson
Canada, and NSERC Canada Research Chairs program. We also wish to honor the memory of our co-author, Han, whose remarkable dedication and valuable contributions greatly enriched this work. Her passion and kindness will be fondly remembered and deeply missed.

\normalem
\bstctlcite{IEEEexample:BSTcontrol}
\bibliographystyle{IEEEtran}
\bibliography{reference}


\end{document}